\documentclass[structabstract]{aa}  

\usepackage{color}
\usepackage{graphicx}
\usepackage{natbib}
\bibpunct{(}{)}{;}{a}{}{,} %
\usepackage{txfonts}
\begin{document}
   \title{Probing the envelopes of massive young stellar objects with diffraction limited mid-infrared imaging\thanks{This paper is based on data obtained using the ESO VLT at the Paranal Observatory with programme 083.C-0795 and the Subaru telescope, which is operated by the National Astronomical Observatory of Japan.}}

   \author{H.E. Wheelwright
          \inst{1,2}
          \and
          W.J. de Wit\inst{3}
          \and         
          R.D. Oudmaijer\inst{2}
          \and 
          M.G. Hoare\inst{2}
         \and
         S.L. Lumsden\inst{2}
         \and
         T. Fujiyoshi\inst{4}
         \and
          J.L. Close\inst{2}}
   
   \institute{ Max-Planck-Institut f\"{u}r Radioastronomie, Auf dem H\"{u}gel 69,
53121 Bonn, Germany\\\email{hwheelwright@mpifr-bonn.mpg.de}
\and School of Physics and Astronomy, University of Leeds, Leeds
     LS2 9JT, UK    
     \and  European Southern Observatory, Alonso de Cordova 3107, Vitacura, Santiago, Chile
   \and Subaru Telescope, National Astronomical Observatory of Japan, 650 North A'ohoku Place, Hilo, HI 96720, USA}

   \date{Received month dd, yyyy; accepted Month dd, yyyy}

   \abstract {Massive stars form whilst they are still embedded in
     dense envelopes. As a result, the roles of rotation, mass loss
     and accretion in massive star formation are not well understood.}
   {This study evaluates the source of the $Q-$band,
     {{$\lambda_{\rm{c}}=19.5~\mu$m}}, emission of massive young
     stellar objects (MYSOs). This allows us to determine the relative
     importance of rotation and outflow activity in shaping the
     circumstellar environments of MYSOs on 1000~AU scales. } {We
     obtained diffraction limited mid-infrared images of a sample of
     20 MYSOs using the VLT/VISIR and Subaru/COMICS instruments. For
     these 8~m class telescopes and the sample selected, the
     diffraction limit, $\sim$0.6'', corresponds to approximately
     1000~AU. We compare the images and the spectral energy
     distributions (SEDs) observed to a 2D, axis-symmetric dust
     radiative transfer model that reproduces VLTI/MIDI observations
     of the MYSO W33A. We vary the inclination, mass infall rate, and
     outflow opening angle to simultaneously recreate the behaviour of
     the sample of MYSOs in the spatial and spectral domains.}  {The
     mid-IR emission of 70 percent of the MYSOs is spatially
     resolved. In the majority of cases, the spatial extent of their
     emission and their SEDs can be reproduced by the W33A model
     featuring an in-falling, rotating dusty envelope with outflow
     cavities. There is independent evidence that most of the sources
     which are not fit by the model are associated with
     ultracompact\,\ion{H}{ii} regions and are thus more evolved.}
   {We find that, in general, the diverse $\sim$20~$\mu$m morphology
     of MYSOs can be attributed to warm dust in the walls of outflow
     cavities seen at different inclinations. This implies that the
     warm dust in the outflow cavity walls dominates the $Q-$band
     emission of MYSOs. In turn, this emphasises that outflows are an
     ubiquitous feature of massive star formation.}

\titlerunning{Diffraction limited MIR imaging of MYSOs}
\authorrunning{H.E. Wheelwright et al.}
\maketitle
%

\section{Introduction}
Massive stars, $M \ga 8~M_{\odot}$, can play a dominant role in the
evolution of their host galaxies via feedback phenomena such as
ionizing radiation, strong stellar winds and supernova
explosions. These processes exert a large influence on the outcome of
subsequent star formation (SF). Massive SF can therefore be considered
as an integral part of the overall SF process and a key ingredient for
a complete description of it. However, our comprehension of many
aspects of massive SF is still unsatisfactory. For example, it is
still not known whether massive SF is intrinsically different to the
generally much better understood process of SF at low masses or not
\citep[see e.g.][]{ZinneckerandYorke2007}.

\smallskip

The uncertainties regarding massive SF are in part the result of the
short Kelvin-Helmholtz (KH) timescale associated with massive stars:
they contract rapidly to sustain a high luminosity while still
acquiring their mass. If the accretion proceeds spherically, the
ensuing radiation pressure could limit the final stellar mass to
approximately 40~$M_{\odot}$
\citep[][]{Larson1971,Kahn1974,Wolfire1987}. This led to the
suggestion that massive SF is intrinsically different to the low mass
case. However, recent radiation-hydrodynamical simulations clearly
demonstrate that the radiation is actually channelled through
optically thin polar cavities while accretion continues through an
equatorial disc
\citep[][]{YorkeandSonnhalter2002,Krumholz2009,Kuiper2010}. Therefore,
deviations from spherical symmetry, which are suggested by the
spectral energy distributions of young massive stars
\citep[][]{Harvey1977,Guertler1991}, remove the limit imposed by
spherical accretion. Nonetheless, confirming that massive stars build
up their final mass via disc accretion by direct observations is
challenging. Massive young stellar objects (MYSOs) are generally
located at kpc distances and their AU-scale circumstellar environment
remains unresolved with single dish telescopes at most wavelength
regions \citep[see e.g.][]{Cesaroni2007}.

\smallskip

Long-baseline infrared interferometry is one of the few techniques
which offers the milli-arcsecond (mas) resolution required to probe
for MYSO discs \citep[see
e.g.][]{Follert2010,Vehoff2010,deWit2011,Grellmann2011}. In
particular, an exemplary result is delivered by \citet{Kraus2010}
using the Very Large Telescope Interferometer (VLTI) and the
near-infrared beam-combiner AMBER. These authors reconstruct a
$K$-band synthesis image of a 20$M_{\odot}$ MYSO with an angular
resolution of 2.4~mas. The resultant image shows hot material in a
disc-like geometry of approximately 20~AU in size. As a result, a
consensus is beginning to emerge that the disc accretion scenario can
account for the formation of stars up to at least $\sim
30\,M_{\odot}$. This implies that the MYSO circumstellar environment
deviates significantly from spherical symmetry. In this scenario,
rotation, accretion discs and outflow activity play a crucial role in
providing the asymmetric environment required to facilitate accretion
\citep[see e.g.][]{Beuther2002}.

\smallskip 

The combination of a high stellar luminosity with accretion and
outflow activity results in the circumstellar environment of MYSOs
being shaped by several forces. Determining the relative importance of
these phenomena in sculpting the circumstellar environments of MYSOs
requires spatially resolved observations. In particular, the
mid-infrared (MIR) wavelength region provides a wealth of information
on the geometry of the accretion environment. Furthermore, this
information can be accessed by both interferometric techniques and
diffraction limited single-dish imaging with 8m class telescopes. MIR
interferometric observations reveal that the $N$-band emission of
MYSOs on scales of 50 to 100~AU is dominated by warm dust located in
the envelope \citep[][]{Dewit2007,Linz2009,Vehoff2010}. It is expected
that such envelopes will contain outflow cavities as there are many
detections of outflow activity associated with MYSOs
(e.g. \object{G35.20-0.74} in De Buizer 2006\nocite{DeBuizer2006};
\object{IRAS 20126+4104} in De Buizer 2007\nocite{DeBuizer2007};
\object{Cep\,A HW2} in de Wit et al. 2009\nocite{deWit2009}). In
\citet[][]{deWit2010}, we argue that, specifically, the warm dust
responsible for the $N-$band emission of W33A is located in the
envelope close to the walls of the cavities evacuated by the outflow.

\smallskip

The aim of this paper is twofold. We aim to assess the source of the
MIR emission of a sample of MYSOs and thus test the premise that their
MIR emission is dominated by warm dust located in outflow cavity
walls. This will then allow us to determine whether outflows play an
important role in shaping the environments of MYSOs ({{as opposed to,
    for example, rotation}}). We address these goals by comparing
spatially resolved images of a sample (20 objects) to appropriate
models. Our approach is similar to the work presented previously in
\citet[][hereafter DW09]{deWit2009} with the addition of 2D
modelling. We present MIR VLT images observed at $20\,\mu$m for a
sample of MYSOs. We then compare the spatially resolved images to
two-dimensional, axis-symmetric dust radiative transfer models that
feature rotating, collapsing envelopes and outflow cavities \citep[the
models of][]{W22003,W12003}. The paper is structured as follows. We
describe the sample selection, the observations and data reduction
procedures in Sect. \ref{obs}. The MIR images are described in
Sect. \ref{data} and the radiative transfer model and its use are
detailed in Sect. \ref{model}. The results of the modelling are
presented in Sect. \ref{mod_res} and discussed in
Sect. \ref{disc}. Finally, our conclusions are given in
Sect. \ref{conc}.

\section{Selection, observations and data reduction}
\label{obs}

We aim to perform our analysis on a representative sample of MYSOs, in
order to draw general conclusions. Selecting such a sample is not
trivial. Initial attempts employed $IRAS$ data \citep[see
e.g.][]{Palla1991,Molinari1996,Sridharan2002}. As a result, such
catalogues suffered from source confusion due to the large beam of
$IRAS$. To rectify this problem and generate a Galaxy-wide, unbiased,
MYSO sample, we have conducted a survey to detect and characterise
MYSOs: the RMS survey \citep[][]{Lumsden2002,James-RMS}. The RMS is
based on data of the $MSX$ survey \citep[][]{Egan2003}. $MSX$ data
offers a significant improvement over $IRAS$ data in terms of
resolution (e.g. arcsecond rather than arcminute resolution), and thus
enables the selection of a representative sample of MIR bright MYSOs.

\smallskip

From the RMS database\footnote{http://www.ast.leeds.ac.uk/RMS/} we
selected objects with distance estimates within 3~kpc\footnote{Some of
  distances estimates have since been revised to larger values.} and
with 21\,$\mu$m $MSX$ fluxes larger than 30\,Jy. The flux limit was
imposed in order to obtain a decent signal-to-noise ratio in the wings
of the resolved profiles. In total, we observed 19 objects at
20~$\mu$m using the VLT Imager and Spectrometer for the Mid Infrared
\citep[VISIR,][]{VISIR} mounted at the Cassegrain focus of UT3 of the
VLT (see Table\ref{t1}). Observations were conducted using the imaging
mode of the instrument and the Q3 filter which has a central
wavelength of 19.5~$\mu$m and a half-band-width of 0.4~$\mu$m. During
the observations, VISIR was equipped with a DRS 256$\times$256 Si:As
detector with an angular pixel size of 0.127\arcsec. This
configuration enabled us to obtain oversampled, diffraction limited
images, which with 8~m class telescopes results in an angular
resolution of approximately 0.6\arcsec.

\smallskip

In this paper, we also present the imaging observations and analysis
of a key MYSO object \object{W33A} (\object{G012.9090-00.2607}). It
was observed at the slightly longer wavelength of 24.5~$\mu$m with the
COMICS instrument \citep[Cooled Mid Infrared Camera and
Spectrometer][]{Kataza2000} mounted on the Cassegrain focus of the
Subaru telescope. We employed the imaging facility of COMICS and used
a filter centred at 24.5~$\mu$m (see DW09 for filter response
functions). During the observations, COMICS was equipped with a
320$\times$240 Si:As IBC detector which provides oversampled
diffraction limited images with a pixel size of 0.13\arcsec. A log of
all the observations is presented in Table \ref{t1}.

\smallskip

\begin{center}
  \begin{figure}[t]
    \begin{center}
      \includegraphics[width=0.4\textwidth]{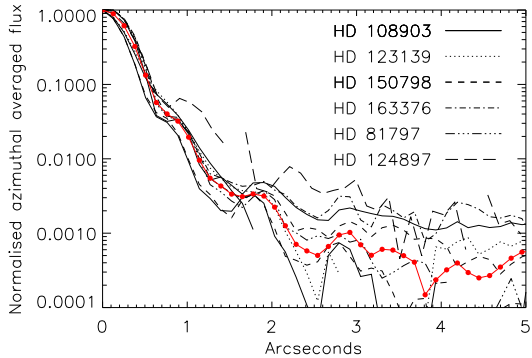} 
      \caption{The azimuthally averaged intensity profiles of the PSF standard stars. HD\,124897 is 
        the PSF standard taken with Subaru/COMICS. The solid line with square points is the average VISIR PSF.\label{psfs}}
    \end{center}
  \end{figure}
\end{center}

For both the VISIR and COMICS observations, standard stars were
observed to provide a reference point-spread-function (PSF). The
standard stars selected are MIR bright,$\sim$100~Jy at 25~$\mu$m ,
isolated sources which are expected to be unresolved. Data reduction,
consisting of shifting and adding nodded and chopped images, was
conducted using routines written in {\sc{idl}}. The resultant images
were not astrometrically corrected. Therefore, the images are
presented in terms of distance from the peak of the intensity
distribution of the target source. The azimuthally averaged intensity
profiles of the standards are displayed in Fig. \ref{psfs}. The radial
profiles of the VISIR standard stars are generally consistent until a
level of approximately 0.001 times the peak flux and appear to behave
as point sources. The standard star of the COMICS data (HD 124897) is
of lower signal-to-noise ratio but is similar out to 1 percent of the
peak flux to the standards presented in DW09.

\begin{table*}
  \caption{The log of the observations.\label{t1}}
  \begin{small}
    \begin{tabular}{llllccp{0.5cm} p{0.25cm} p{0.25cm} p{0.25cm}}
      \hline
      \vspace*{-2.5mm}
      \rule{0pt}{0.5mm}\\
      Object & RA &DEC(J2000)  & date & Integration &  $\overline{\mathrm{AM}}$ & \multicolumn{1}{c}{S/N}& \multicolumn{1}{c}{$d$} & \multicolumn{1}{c}{$L$}&$F_{\rm 21\mu m}$ \\
      & (h:m:s) & (\degr:\arcmin:\arcsec)         &          &  (s) & & & \multicolumn{1}{c}{(kpc)} & \multicolumn{1}{c}{($\mathrm{10^3}L_{\odot}$)}&(Jy)\\
      \hline
      \hline
      {\bf{MYSOs:}}\\
      {\bf{VISIR}}\\
      G263.7434+00.1161   &  08 48 48.64 &$-$43 32 29.1 & 09/05/11              & 760  	          &1.1       & 900      &  2.0     & 4.1 & 93\\
      G263.7759$-$00.4281 &  08 46 34.84 &$-$43 54 29.9 & 09/05/11+05/12  & 1500           &1.1       & 150       &  1.5     & 4.0 & 65\\
      G265.1438+01.4548   &  08 59 27.40 &$-$43 45 03.7 & 09/06/04+05/05   & 1833           &1.4       & 150      &  2.5     & 8.6 & 42\\
      G268.3957$-$00.4842 &  09 03 25.08 &$-$47 28 27.6 & 09/06/05+06/22   & $2\times$ 1305  & 1.6/1.7& 250 &  1.6     & 3.7 & 50\\ 
      G269.1586$-$01.1383 &  09 03 31.76 &$-$48 28 45.5 & 09/05/12              & 700             &1.2       & 25        &  2.7     & 7.3 & 98\\
      G310.0135+00.3892   &  13 51 37.85 &$-$61 39 07.5 & 09/05/26               & 120             &1.4       & 250      &  3.3     & 57 & 259\\
      G314.3197+00.1125   &  14 26 26.28& $-$60 38 31.5 & 09/05/26               & 2000           &1.3       & 600      &  3.7     & 13 & 58\\
      G318.0489+00.0854   &  14 53 42.99 &$-$59 08 56.5 & 09/06/01+06/07   & $2\times$ 1996  &1.2/1.2 & 50    &  ?       & ?  & 41\\
      G318.9480$-$00.1969 &  15 00 55.31 &$-$58 58 52.6 & 09/05/10              & 2000           &1.2       & 650      &  2.6     & 11 & 55\\
      G326.4755+00.6947   &  15 43 18.97 &$-$54 07 35.6 & 09/06/07              & 1851            &1.2       & 650      &  2.8     & 8.8 & 42\\
      G332.2941+02.2799   &  16 05 41.84 &$-$49 11 29.5 & 09/06/07              & 3199   	  &1.2       & 75        &  1.9     & 0.8 & 32\\
      G332.9868$-$00.4871 &  16 20 37.81 &$-$50 43 49.6 & 09/06/07             & 1500            &1.1       & 500      &  3.5     & 26 & 64\\
      G333.1075$-$00.5020 &  16 21 14.22 &$-$50 39 12.6 & 09/06/07             & 1447            &1.4       & 50        &  3.5     & 2.9 & 48\\
      G339.6221$-$00.1209 &  16 46 05.99 &$-$45 36 43.9 & 09/06/07             & 2225            &1.4       & 150      &  13.1   & 520 & 39\\
      G341.1281$-$00.3466 &  16 52 33.19 &$-$44 36 10.8 & 09/06/08             & 2423  	  &1.1       & 400      &  3.5     & 6.2 &37\\
      G343.5024$-$00.0145 &  16 59 20.90 &$-$42 32 38.4 & 09/06/07             & 400              &1.4       & 25        &  3.0     & 18 & 131\\
      G343.5213$-$00.5171 &  17 01 34.04 &$-$42 50 19.7 & 09/05/10             & 1497            &1.1       & 75        &  3.0     & 13 & 47\\
      G345.0061+01.7944   &  16 56 46.37 &$-$40 14 26.7 & 09/05/10              & 280             &1.0       & 100      &  1.8      & 8.7 & 154\\
      G349.7215+00.1203   &  17 18 11.11& $-$37 28 23.4 & 09/06/07              & 3520           &1.6       & 200      &  24.4    & 310 & 31\\
      {\bf{COMICS}}\\
      G012.9090$-$00.2607 & 18:14:39.56 &$-$17:52:02.3 & 09/08/30 & 1405 & 1.3 & 200& 3.8& 54& 144\\
      \hline
      {\bf{Standards:}}\\
      {\bf{VISIR}}\\
      HD 108903 & 12 31 09.96 &$-$57 06 47.6  & 09/05/28+06/07+06/08 & 60/60     & 2.0/1.4/1.4 & 200/400/700& \\
      HD 123139 & 14 06 40.95 &$-$36 22 11.8  & 09/06/23             & 2500      & 1.0         & 350&\\
      HD 150798 & 16 48 39.90 &$-$69 01 39.8  & 09/05/10+05/26       & 3600/3600 & 1.4/1.4     & 1800/600&\\
      HD 163376 & 17 57 47.80 &$-$41 42 58.7  & 09/05/26             & 3600      & 1.1         & 150&\\
      HD 81797  & 09 27 35.24 &$-$08 39 31.0  & 09/05/12             & 2500      & 1.2         & 850&\\
      {\bf{COMICS}}\\
      HD 124897  & 14 15 39.67 &+19 10 56.7 & 09/08/30& 856& 1.9 & 50 \\
      \hline
    \end{tabular}
    \tablefoot{The signal to noise ratio is defined as the ratio of the central peak to the root-mean-squared (rms) background noise. 
      The distances and luminosities are taken from the RMS database.}
  \end{small}
\end{table*}

\section{Observational results}

\label{data}

\begin{center}
  \begin{figure*}[!h]
    \begin{center}
      \begin{tabular}{l l l}
        \includegraphics[width=0.3\textwidth]{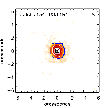} & 
        \includegraphics[width=0.3\textwidth]{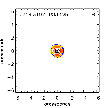}&
        \includegraphics[width=0.3\textwidth]{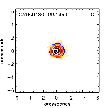}\\
        \includegraphics[width=0.3\textwidth]{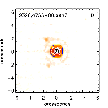}&
        \includegraphics[width=0.3\textwidth]{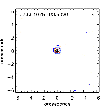}&
        \includegraphics[width=0.3\textwidth]{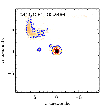}\\
      \end{tabular}
      \caption{Unresolved sources\label{unres_im}. The images are
        scaled logarithmically. North is to the top of the page and
        East is to the Left. {{The contours typically represent 1, 5,
            25 and 75 percent of the peak flux.}}}
    \end{center}
  \end{figure*}
\end{center}

We assessed the MIR images of our target sources by comparing their
azimuthally averaged intensity profiles to the instrumental PSF
determined from the standard stars. We find that the majority of the
MYSOs observed, 14 of 20, are spatially resolved. The images of the
target MYSOs are presented in Figs. \ref{unres_im} and \ref{res_im},
which contain the unresolved and resolved MYSOs respectively.

\smallskip

The resolved MYSOs generally appear as a single source within the
VISIR field of view, $30\arcsec\times30\arcsec$. Most display an
approximately circular symmetric morphology at the resolution of the
observations ($\sim$0.6\arcsec), although some show clear signs of
structure (e.g G263.7759$-$00.4281, Fig. \ref{res_im}: panel
A). Several objects are associated with additional, distinct sources
of localised emission, which often have a cometary type morphology
(e.g. G269.1586$-$01.1385 \& G349.7215+00.1203, Fig. \ref{res_im}:
panels D \& M). This is suggestive of a H{\sc{ii}} region
\citep[][]{Hoare2007PPV}. Indeed, both the objects mentioned above are
detected at radio wavelengths at coordinates consistent with the
locations of the cometary 20~$\mu$m sources
\citep[see][]{Urquhart2007a}. It is of interest to note that
24.5\,$\mu$m images of regions containing both MYSOs and ultracompact
H{\sc{ii}} regions show a clear morphological difference between the
two types of source, i.e. compact vs extended (DW09).

\smallskip

Approximately one third of the sample are unresolved. Using the
Kolmogorov-Smirnov test, we investigated whether the resolved and
unresolved samples were drawn from intrinsically different
distributions of MYSOs in terms of distance and luminosity. There is
no significant difference between the distance and luminosity
distributions of the two samples, as might be expected given that the
objects are drawn from a single catalogue. However, the angular size
of a centrally heated source of flux is dependent upon both the
luminosity of central object and the distance to it \citep[angular
size $\propto \frac{\sqrt{L}}{d}$, see e.g.][]{Vinkovic2007}. We find
that the resolved objects generally have a higher value of
$\frac{\sqrt{L}}{d}$ than the unresolved objects. More specifically,
the hypothesis that the unresolved sources have the same
$\frac{\sqrt{L}}{d}$ distribution as the resolved sources can be
discarded at a level of 98 percent significance. Thus, we conclude
that resolved sources appear larger than the unresolved sources as the
resolved sources are generally more luminous {\emph{and}} less distant
than their unresolved counterparts.

\smallskip

\begin{center}
  \begin{figure*}[!h]
    \begin{center}
      \begin{tabular}{l l l}
        \includegraphics[width=0.3\textwidth]{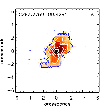} &
        \includegraphics[width=0.3\textwidth]{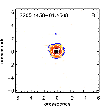} &
        \includegraphics[width=0.3\textwidth]{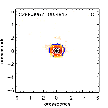} \\
        \includegraphics[width=0.3\textwidth]{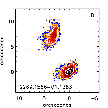}&
        \includegraphics[width=0.3\textwidth]{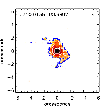}&
        \includegraphics[width=0.3\textwidth]{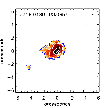}\\
        \includegraphics[width=0.3\textwidth]{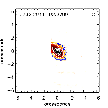}&
        \includegraphics[width=0.3\textwidth]{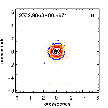}&
        \includegraphics[width=0.3\textwidth]{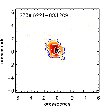}\\
        \includegraphics[width=0.3\textwidth]{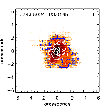}&
        \includegraphics[width=0.3\textwidth]{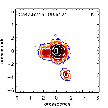}&
        \includegraphics[width=0.3\textwidth]{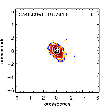}\\
      \end{tabular}
      \caption{Resolved sources\label{res_im}. The images are scaled
        logarithmically. North is to the top of the page and East is to the
        Left. {{The contours typically represent 2, 5, 10, 25 and 75 percent of
            the peak flux.}}}
    \end{center}
  \end{figure*}
\end{center}

\addtocounter{figure}{-1}

\begin{center}
  \begin{figure*}[!h]
    \begin{center}
      \begin{tabular}{ l}
        \includegraphics[width=0.3\textwidth]{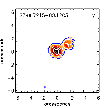} \\
      \end{tabular}
      \caption{continued.}
    \end{center}
  \end{figure*}
\end{center}

\section{Radiative transfer modelling}
\label{model}

To investigate the nature of the circumstellar structure of the
resolved sources, we compare the observed images to 2D axis-symmetric
radiative transfer models appropriate for massive stellar objects
surrounded by in-falling, rotating envelopes. By simultaneously
comparing the resolved MIR images and SEDs to the radiative transfer
calculations, we aim to understand the source of the 20\,$\mu$m MIR
emission of MYSOs. The models and the modelling procedure are
described in detail in the following sections.

\subsection{The radiation transfer code}
We employ the 2D, axis-symmetric dust radiation transfer code of
Whitney et al. \citep[for details see][]{W22003,W12003}. The code
calculates radiation transfer through a dusty structure which consists
of a proto-stellar envelope surrounding a central star. A low density
polar cavity can be inserted into the overall envelope structure. The
code also allows for a dusty circumstellar disc. The density
distribution of the envelope is described by the treatment of a
collapsing, rotating envelope of \citet{Ulrich1976} and
\citet[][]{Terebey1984}. The key parameters determining the
envelope density are the mass infall rate and the centrifugal
radius. The dust sublimation radius is calculated self-consistently
from the stellar luminosity and the dust sublimation temperature. The
shape of the polar cavities can be described in two ways: using a
polynomial function or following streamlines. The streamline is
conical on large scales, while the apparent opening angle of the
polynomial function can be specified (the opening angle at the stellar
surface is 180$\degr$). If included, the dust disc is flared and
follows the $\alpha$ prescription \citep[see][]{Shakura1973} to
account for flux due to accretion. The reader is referred to
\citet{W22003,W12003} for more details on the code and the possible
geometries. We have used this code extensively and the reader is
referred to \citet{deWit2010,deWit2011} for more details on the
application of the code to model MYSOs and their environments.

\subsection{Modelling methodology}

Our previous use of the Whitney et al. code consisted of constructing
a series of dedicated models in order to reproduce the SEDs,
interferometric and auxiliary spatial information of MYSOs. We will
exploit these customised models appropriate for MYSO environments in
our approach here, partially motivated by the lack of such models in
the well-known SED grid \citep[][]{Robitaille2007}. In particular, we
will use a model geometry that provides a successful fit to the source
W33A \citep[][]{deWit2010}. Its defining feature is that the MIR
emission in the $N$-band originates in warm dust located in the
walls of outflow cavities. The model proto-stellar envelope extends
inwards down to the dust sublimation radius and it has low density
outflow cavities as observed in several MIR observations of MYSOs
\citep[see e.g.][]{DeBuizer2007}. We note that the model
does not explicitely include a dust disc (according to the
prescription of the Whitney et al. code). This is motivated by the
aforementioned MIR interferometric observations. Adding a disc
generally results in high visibilities in the MIR whereas MYSOs
typically exhibit low visibilities, indicative of a more extended dust
distribution \citep[see e.g.][]{deWit2010}.

\smallskip

\begin{figure}[t]
  \includegraphics[width=0.5\textwidth]{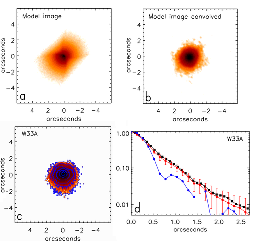} 
  \caption{{\it Panel a:} Logarithmically scaled 24.5~$\mu$m image of
    the standard W33A model. {\it Panel b:} The same image as panel a)
    but convolved with the COMICS instrumental PSF. {\it Panel c:} The
    COMICS image of W33A. {\it Panel d:} The azimuthally averaged
    radial intensity profile at 24.5~$\mu$m of W33A (points with error
    bars) alongside the model radial profile and that of a PSF
    standard (the lower line). The outflow axis of the model is
    aligned with the SE/NW orientation of W33A's
    outflow.\label{test_image}}
\end{figure}

We illustrate the model 24.5\,$\mu$m image for the W33A model in panel
a) of Fig.\,\ref{test_image}. Also shown are the PSF convolved image
(panel b), the COMICS image of W33A and the corresponding radial
intensity profile, panels c) and d) respectively. The shape of the
outflow cavities can clearly be delineated in the model image in panel
a). This demonstrates that the dust located in the outflow
cavity walls is warmed-up and emits a significant fraction of the MIR
flux. W33A is know to drive an outflow in the SE/NW direction
\citep[see][and references therein]{Davies2010}. The orientation of
the outflow is traced by nebulosity in the NIR. While the 24.5~$\mu$m
image of this object is resolved (panel c), it appears relatively
symmetric. However, the image is slightly extended towards the SE, in
the direction of the blue lobe of its outflow. In panel d) we compare
the model's radial intensity profile with the observed one and find
they are essentially identical. We underline that we did not perform a
model fit to the 24.5\,$\mu$m image, we simply used the final W33A
model presented in de Wit et al. (2010) to create the corresponding
image.

Motivated by the good match of the W33A model and image and the
knowledge that the SEDs of MYSOs are known to be similar, we proceeded
to fit the observations of the remaining MYSOs in the following
way. The SEDs of the MYSOs observed were constructed from the
continuum data assembled on the RMS database (which makes use of the
work of various authors) and the methodology of \citet{M2011}. The SED
coverage differs from one source to the other. Generally one can
distinguish between sources with continuum flux measurements up to
100\,$\mu$m and those with sub-mm observations. We note that we do not
fit explicitly the 9.7\,$\mu$m silicate feature, contrary to the
modelling efforts in DW09. This is simply because this information is
not available for the majority of our sources.

\smallskip

To reproduce both the SEDs and 20\,$\mu$m intensity distributions of
the resolved sample, we used the W33A model geometry summarised
above. We only altered parameters that can be expected to vary between
objects: the inclination, the envelope infall rate (which scales the
total dust mass in the envelope) and the opening angle of the outflow
cavity (which may vary with time). We emphasise that we use one basic
model to reproduce the observations. There may be differences between
sources, for example the centrifugal radius and the properties of
circumstellar discs -- if any are present -- may vary from source to
source. The model will not account for such differences. However, it
provides a simple test of the hypothesis that the bulk of the $Q-$band
emission of MYSOs can be attributed to warm dust in outflow cavity
walls. Here we outline the methodology followed in reproducing the
observations.

\smallskip

For a given MYSO, the stellar luminosity was initially set to the
value in Table \ref{t1} and the system inclination {{(the angle
    between the polar axis and the line of sight)}} was estimated from
its MIR image. The stellar luminosity and infall rate were then varied
until the model SED was consistent with that observed. This set the
luminosity and provided first estimates of the other parameters. Once
the SED was satisfactorily reproduced, the inclination and outflow
opening angle were varied in an attempt to recreate the observed
spatial intensity profile. This had an affect on the reproduction of
the SED. Therefore, the infall rate was allowed to vary to enable a
fit to both the SED and the radial profile. The model fitting was done
by hand rather than using a $\chi^2$ minimisation technique as the
computational time required to make a grid of model images for each
object was prohibitively large.

 \smallskip

\begin{center}
  \begin{figure}
    \begin{center}
      \begin{tabular}{l l}
        \includegraphics[width=0.235\textwidth]{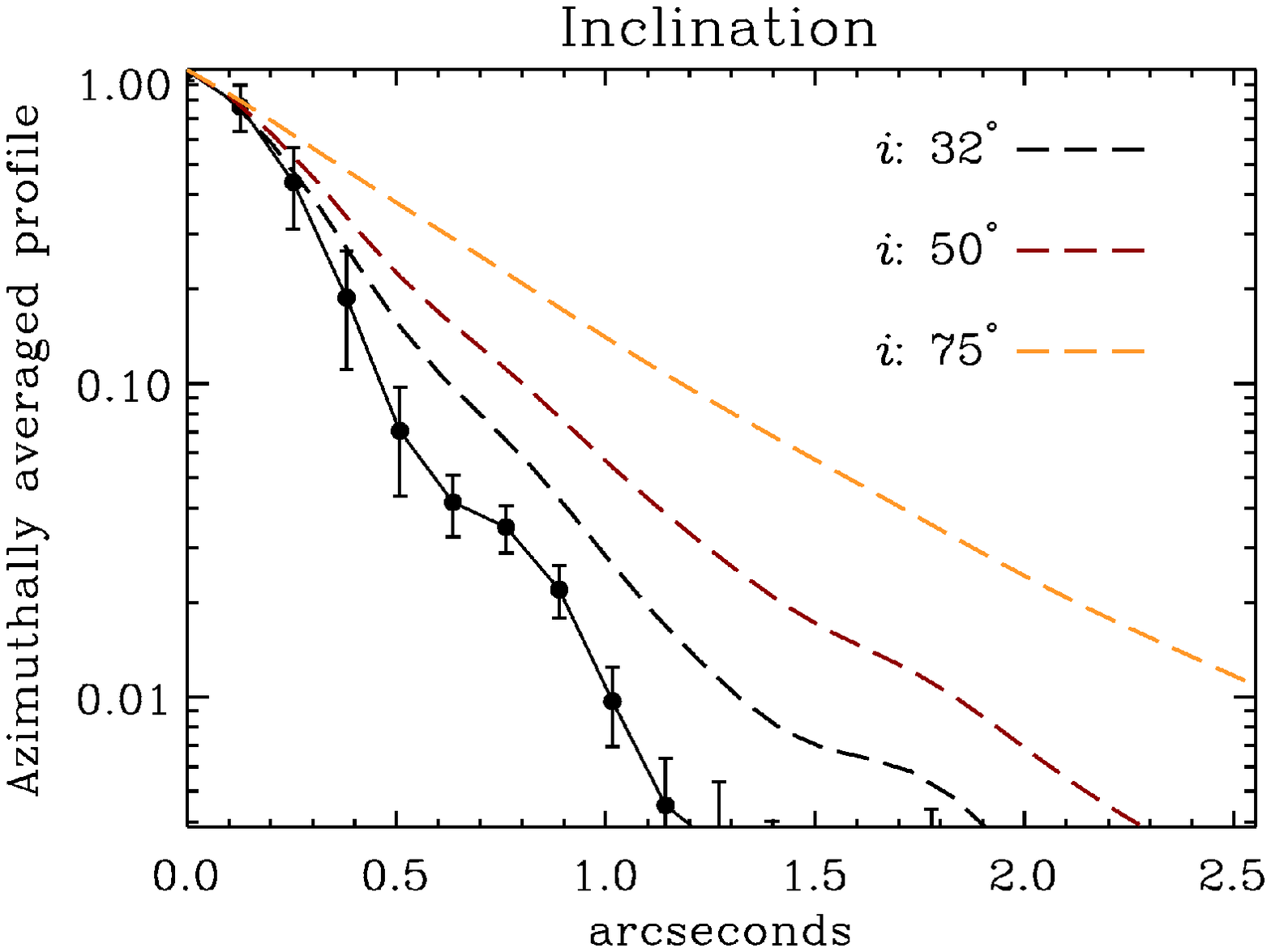} & 
        \includegraphics[width=0.235\textwidth]{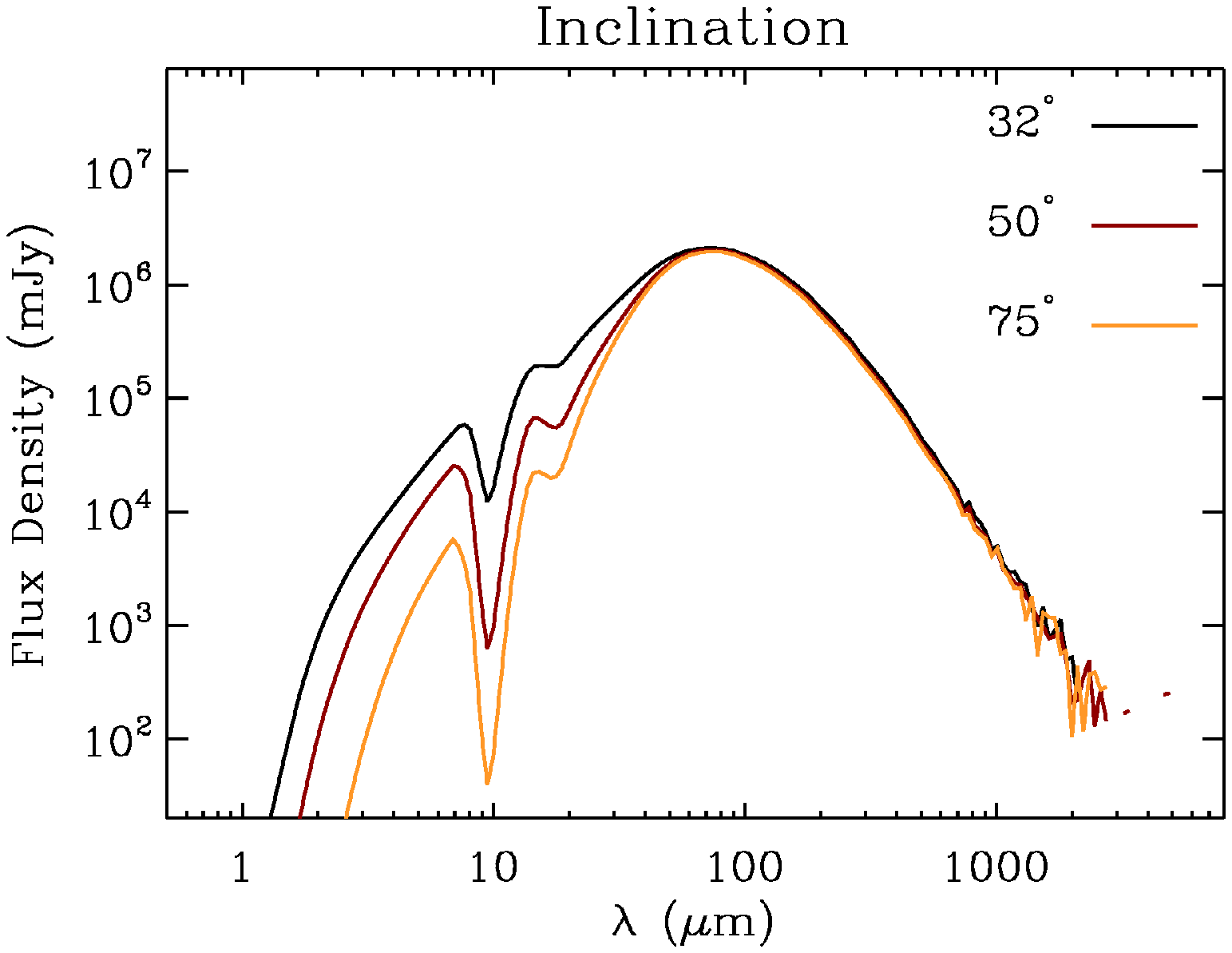}\\
        \includegraphics[width=0.235\textwidth]{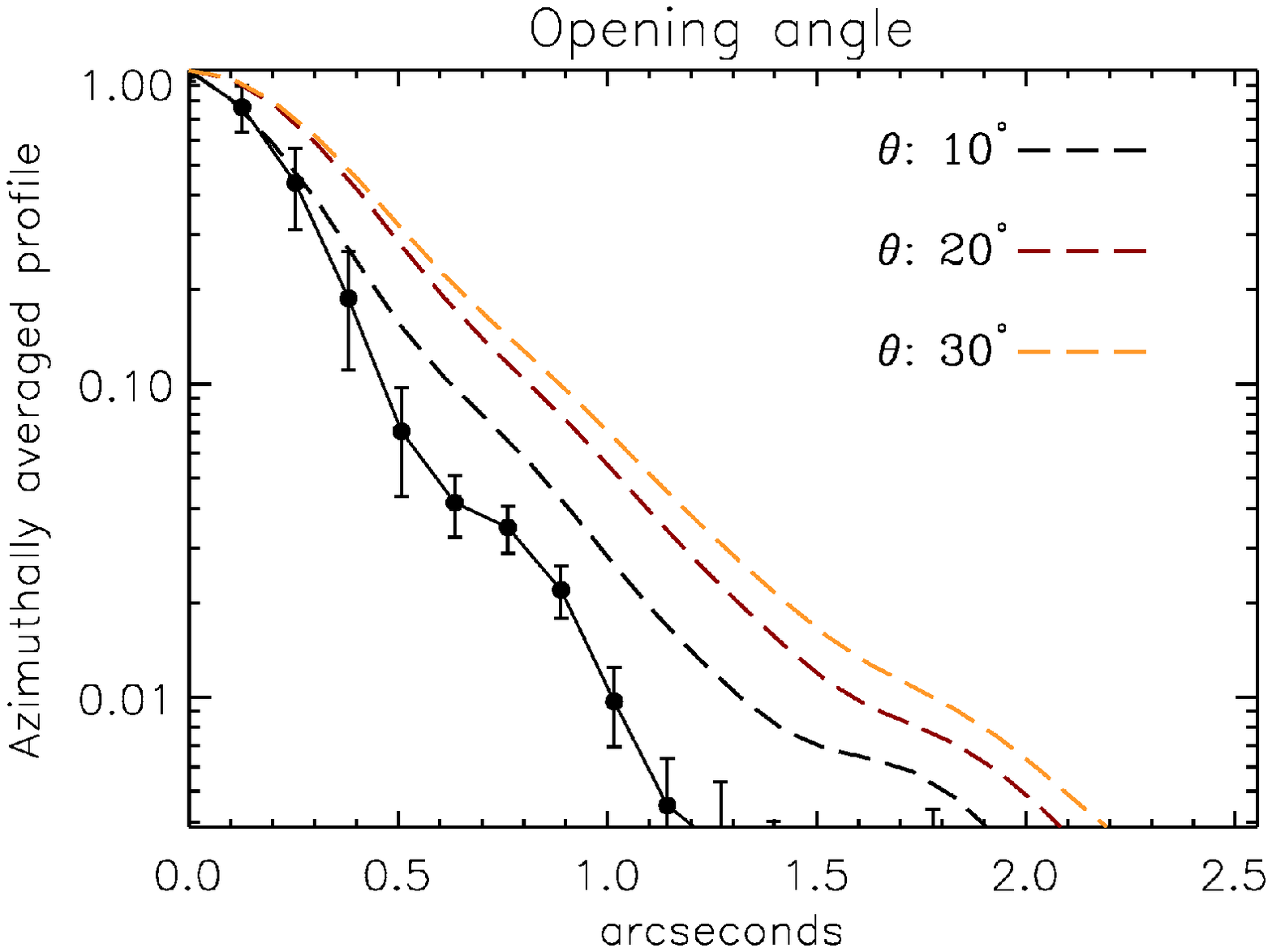}&
        \includegraphics[width=0.235\textwidth]{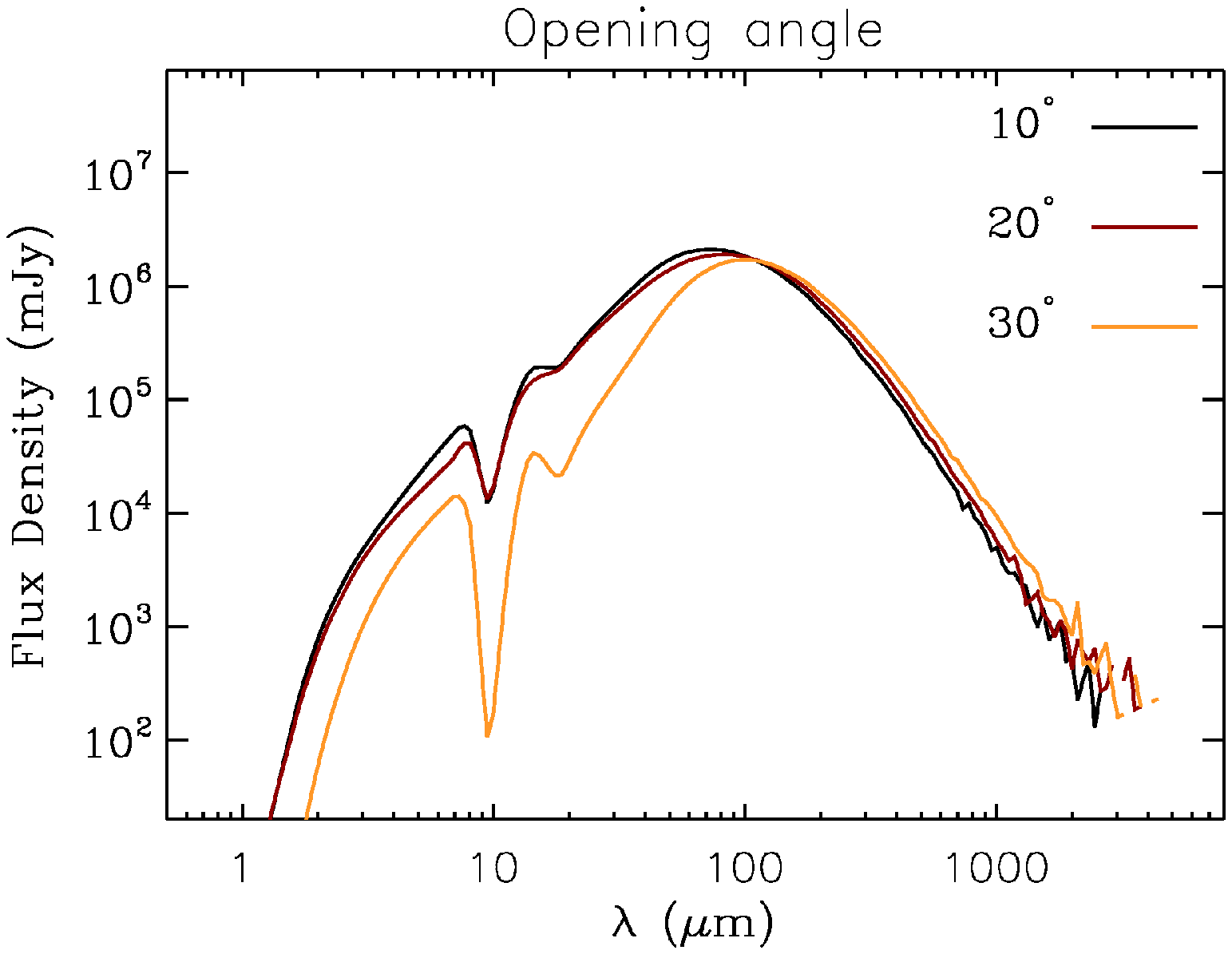}\\
        \includegraphics[width=0.235\textwidth]{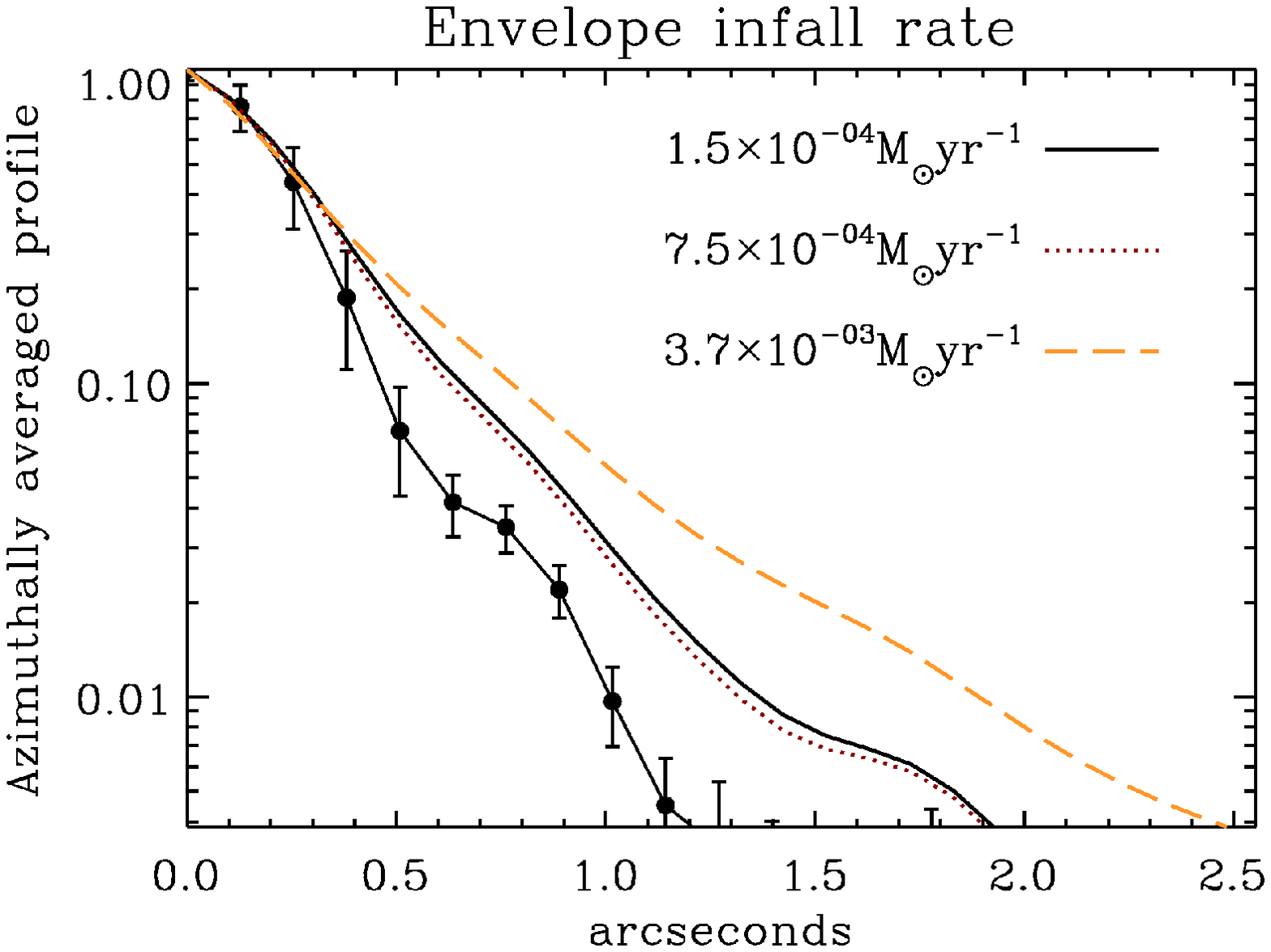}&
        \includegraphics[width=0.235\textwidth]{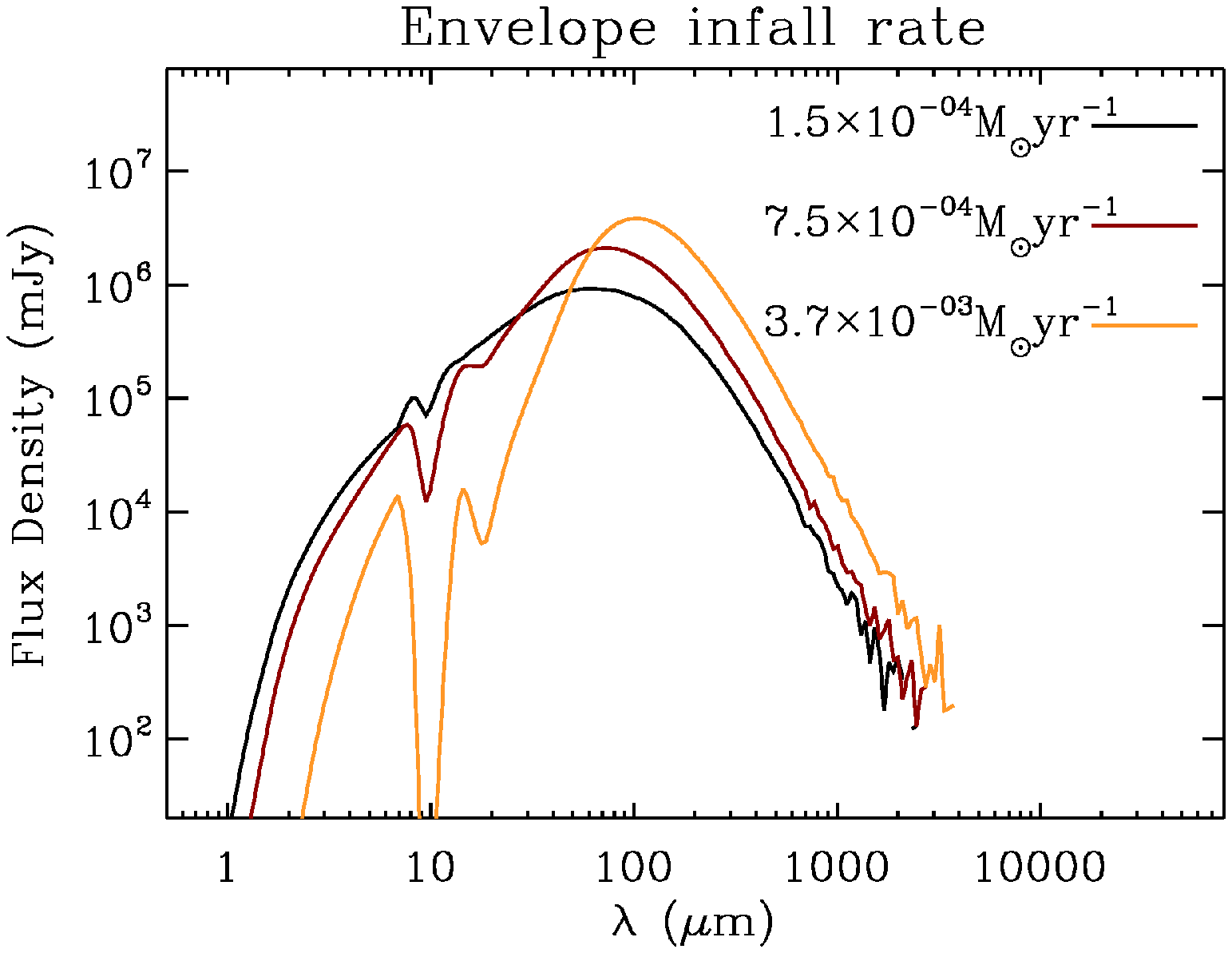} \\
      \end{tabular}
      \caption{An exploration of the parameter space immediately
        surrounding the model of G310.0135+00.3892. The plots show the
        effect of varying the free parameters (the inclination,
        opening angle and infall rate). For comparison, a PSF standard
        is also shown in the intensity distribution
        figures.\label{vary}}
    \end{center}
  \end{figure}
\end{center}

 We demonstrate the effect of varying the free parameters in
 Fig. \ref{vary}. In general, the more inclined models appear more
 extended. This might be expected if much of the MIR emission
 originates in outflow cavity walls. At large inclinations, the projected
 area of the outflow cavity walls is larger and thus the resultant flux
 distribution is more extended. The cavity opening angle has a similar
 effect on the intensity distribution. This is because increasing the
 opening angle increases the distance at which the majority of the
 cavity wall is exposed to direct irradiation from the central
 object. This also increases the projected area of the outflow
 cavities and the extension seen in the azimuthally averaged intensity
 distribution. Since both parameters affect the observed extension,
 they are slightly degenerate. However, the two parameters have
 different effects on the gradient of the radial profile (see
 Fig. \ref{vary}).

\smallskip

\begin{center}
  \begin{figure*}[!h]
    \begin{center}
      \begin{tabular}{l l l}
        \includegraphics[width=0.3\textwidth]{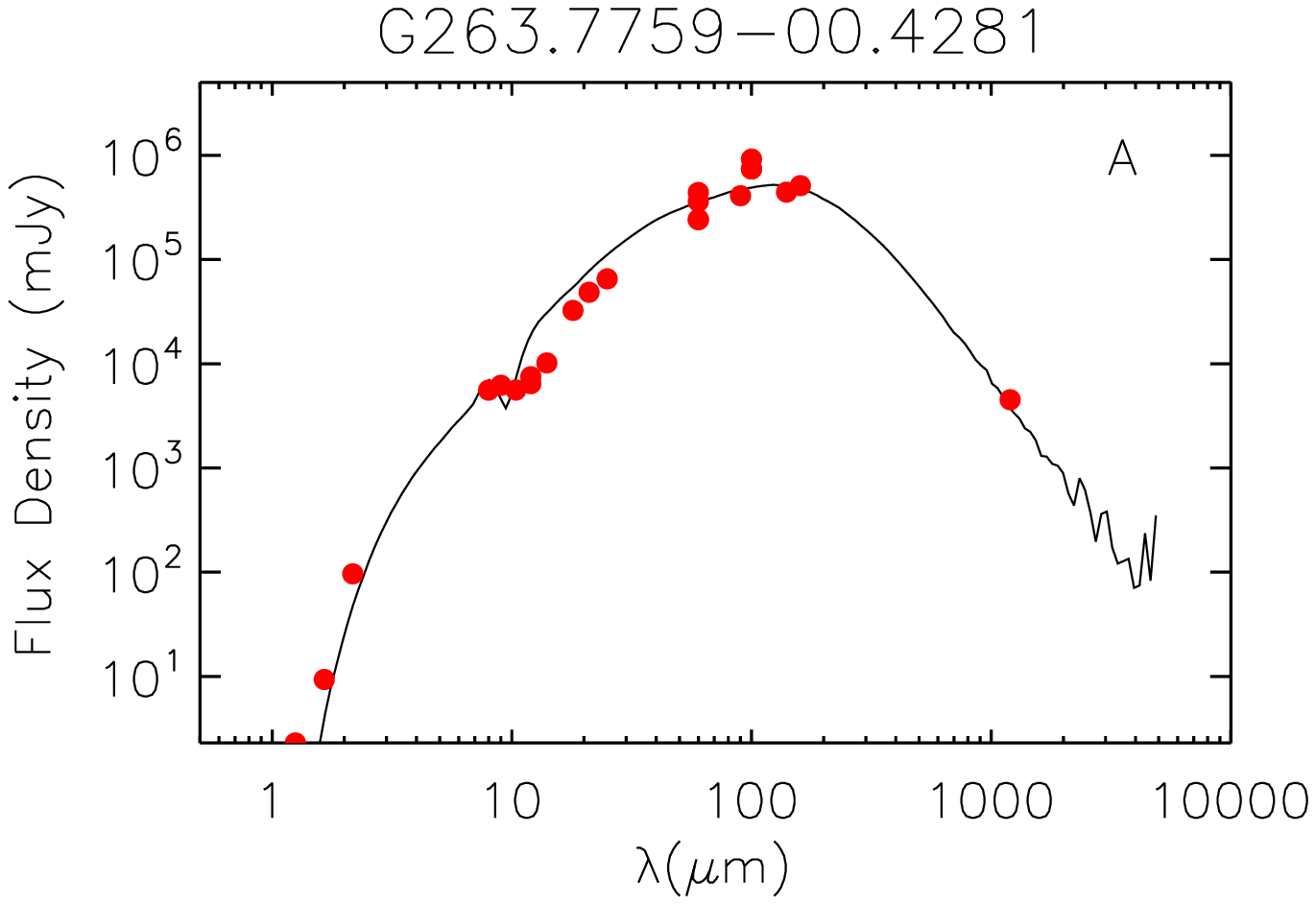} &
        \includegraphics[width=0.3\textwidth]{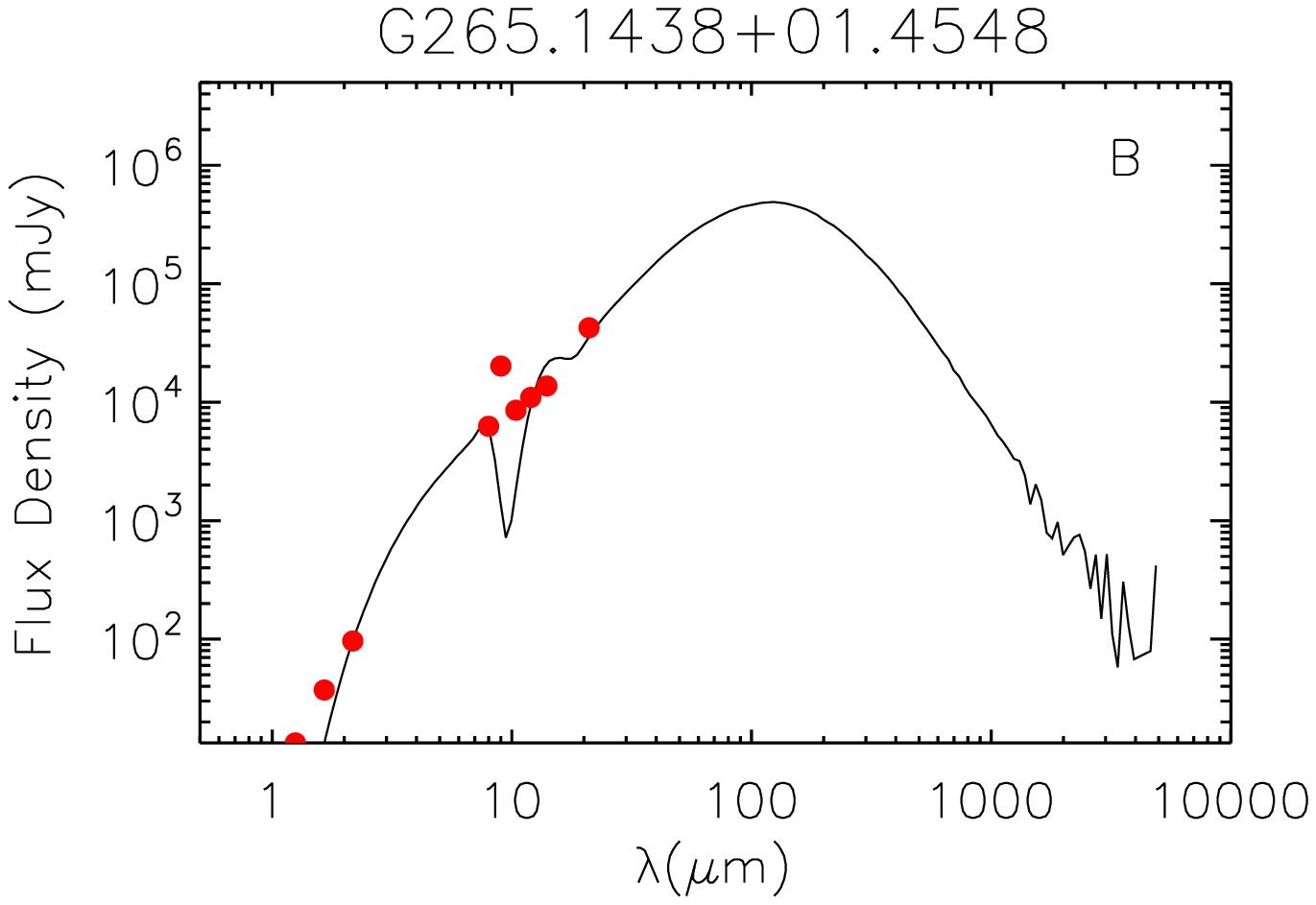} &
        \includegraphics[width=0.3\textwidth]{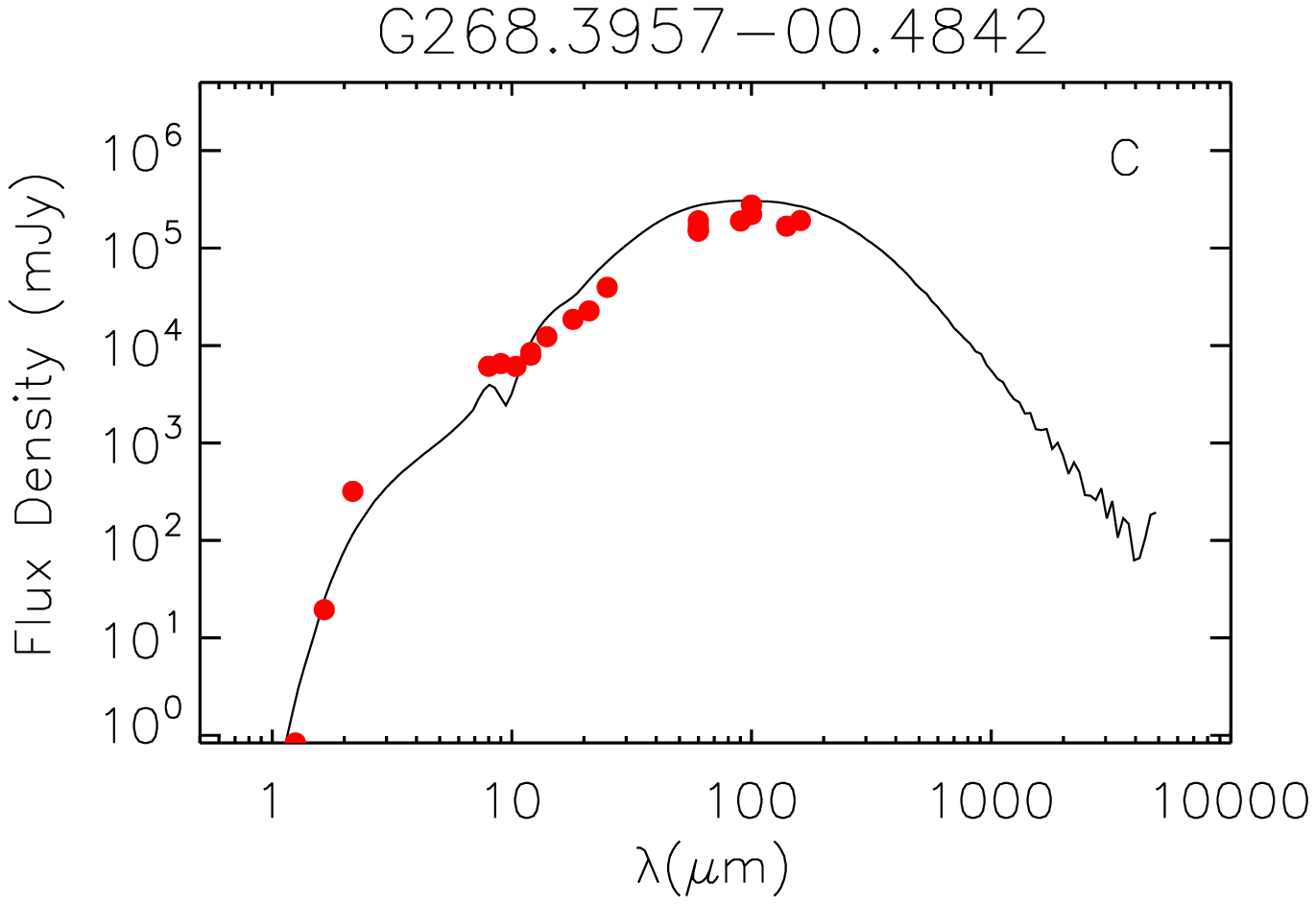} \\
        \includegraphics[width=0.3\textwidth]{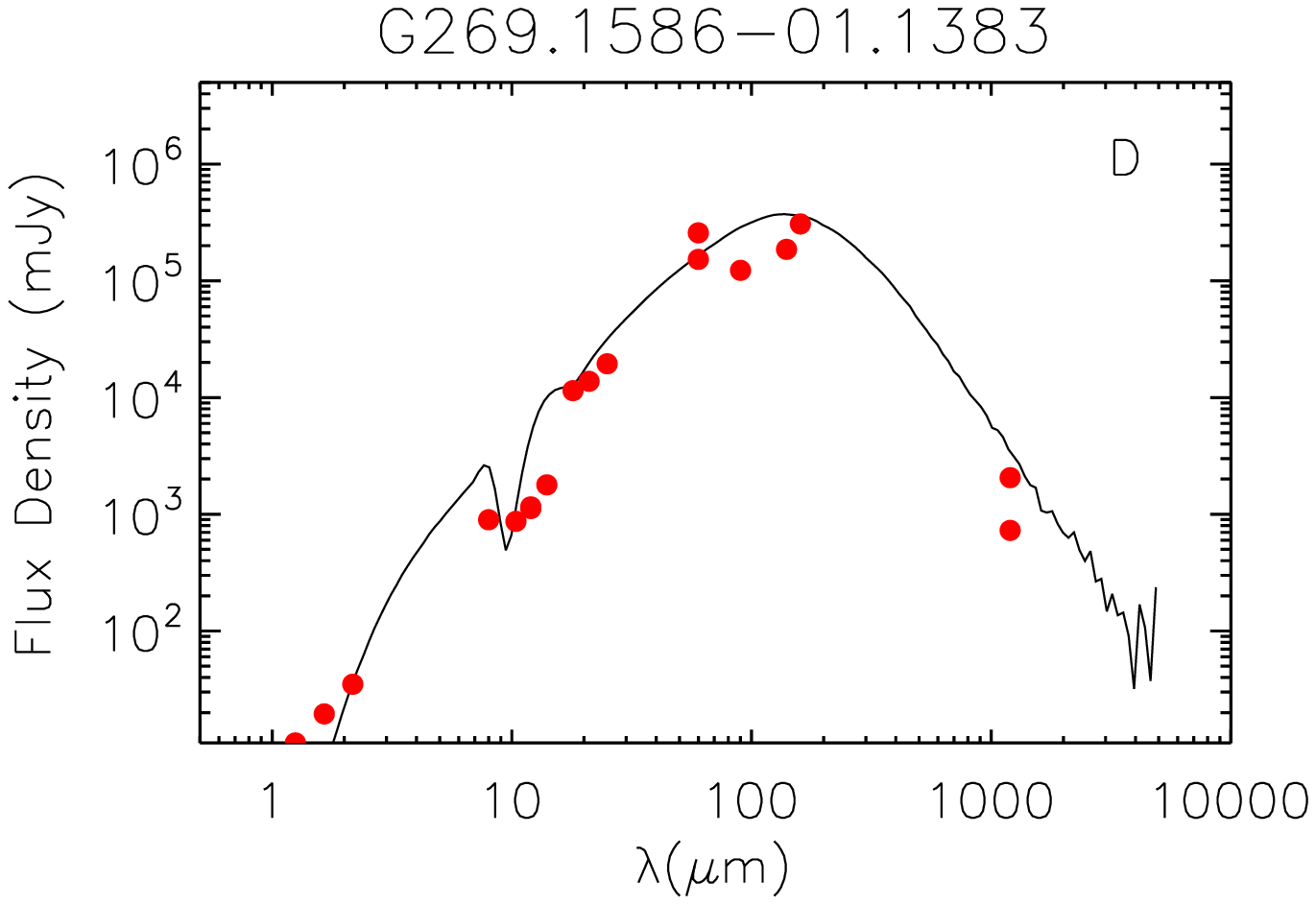}&
        \includegraphics[width=0.3\textwidth]{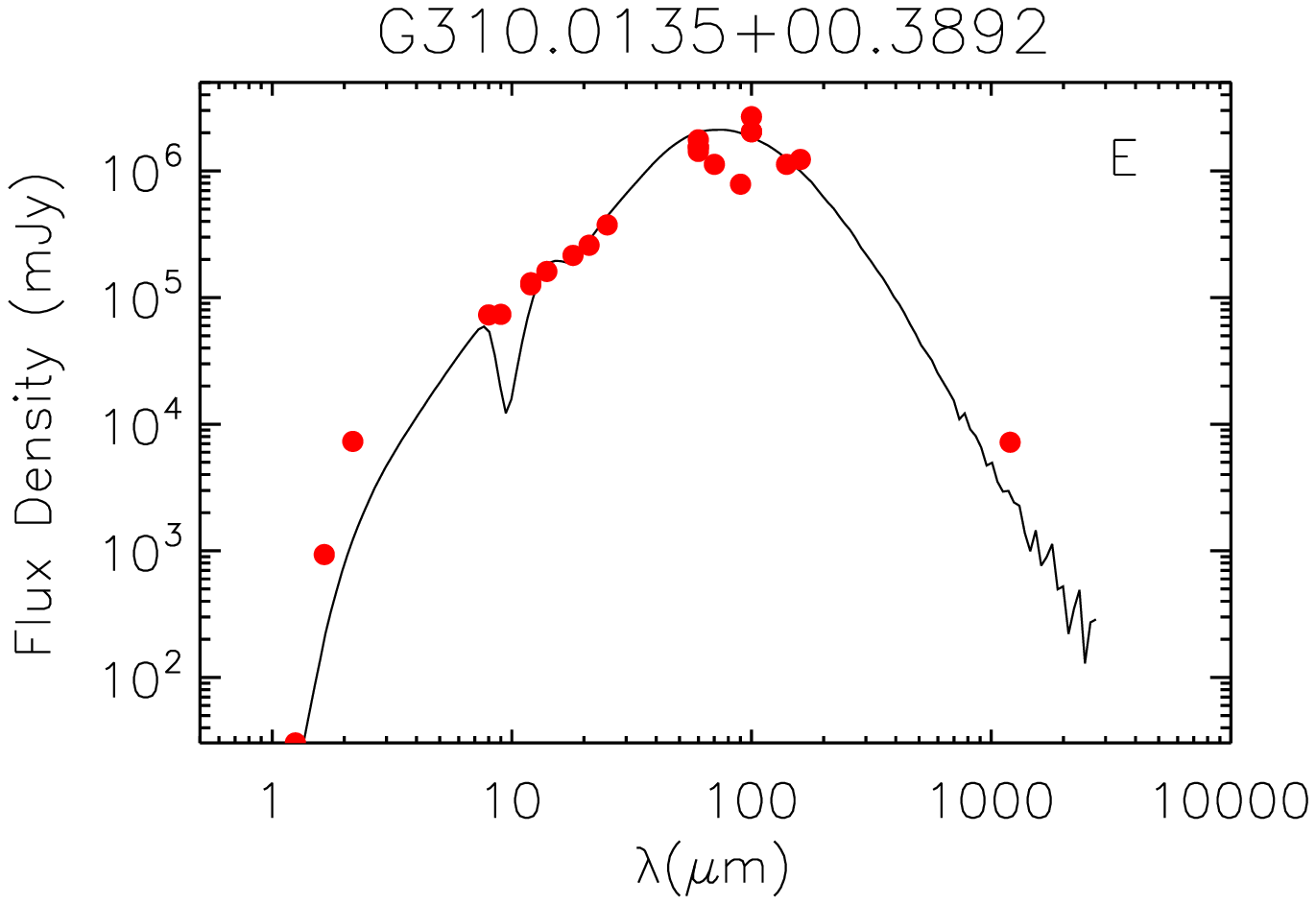}&
        \includegraphics[width=0.3\textwidth]{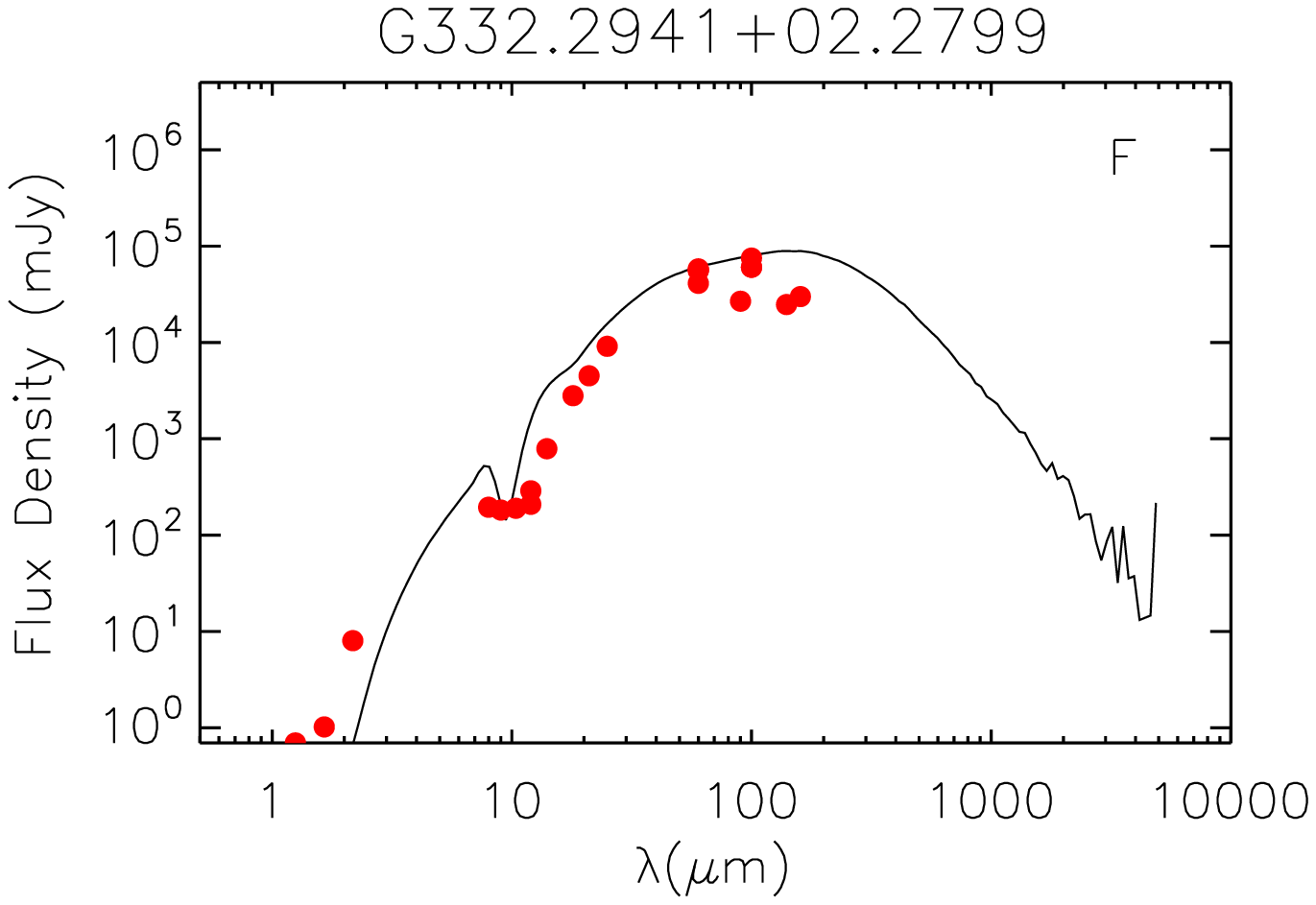}\\
        \includegraphics[width=0.3\textwidth]{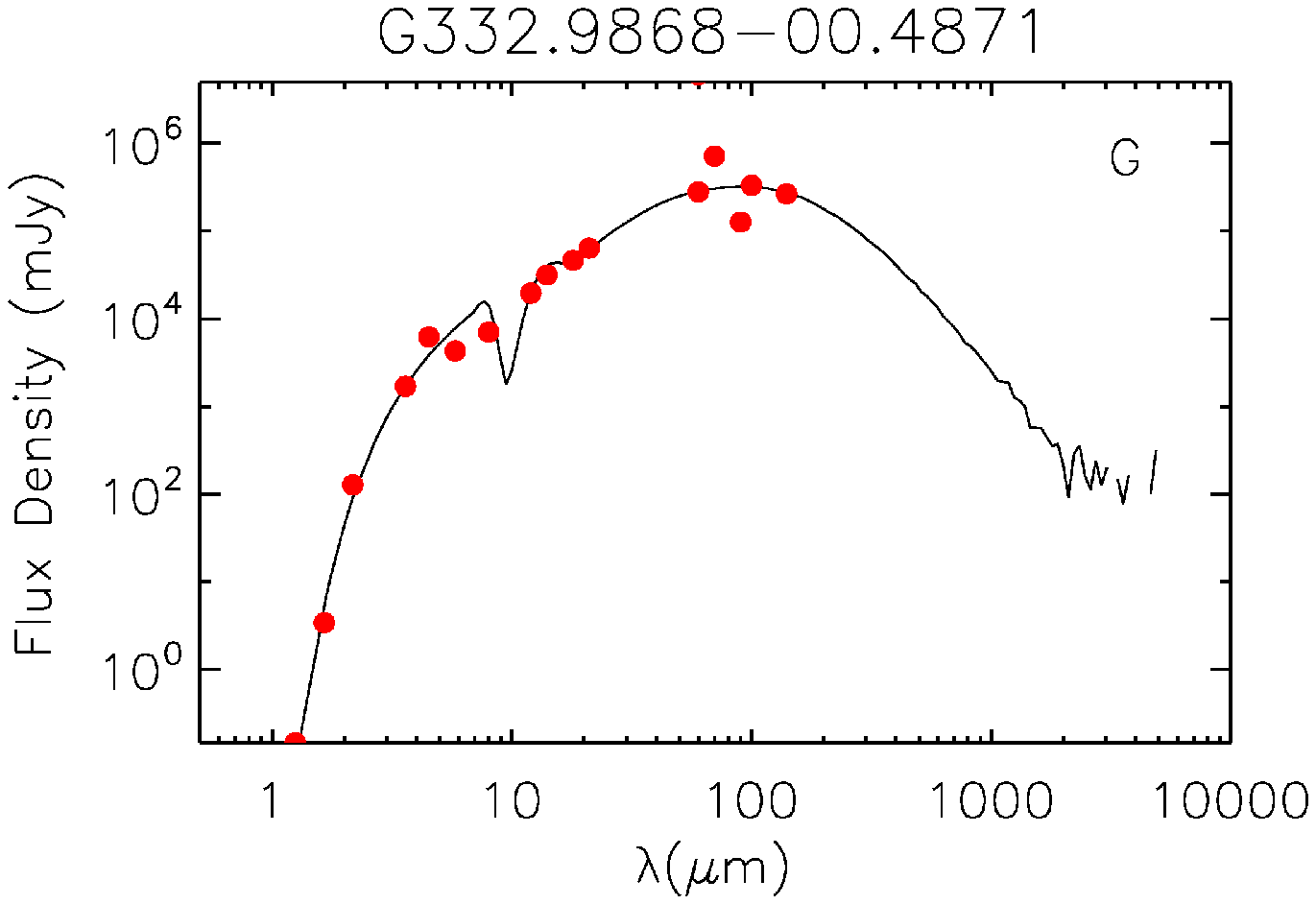}&
        \includegraphics[width=0.3\textwidth]{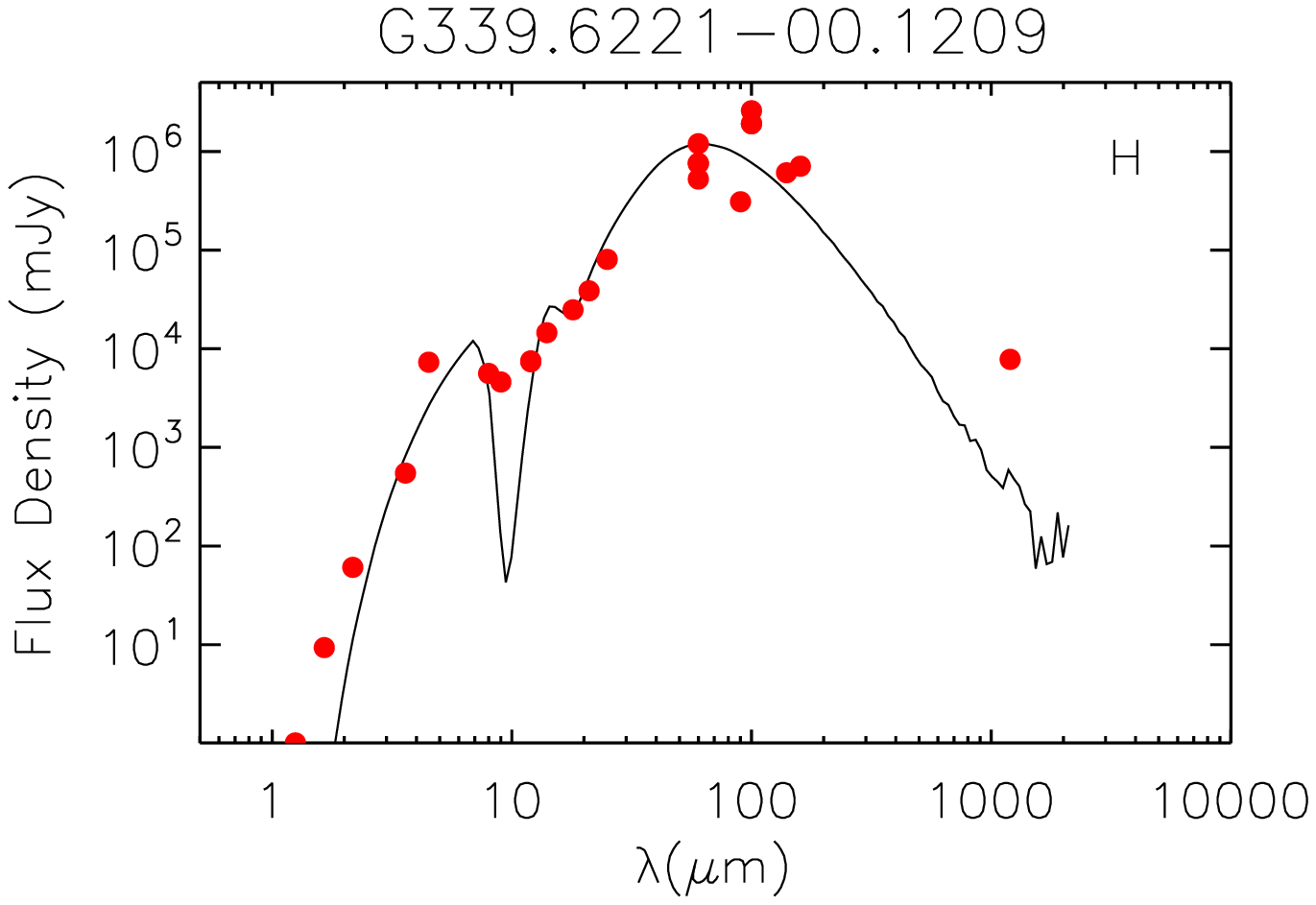}&
        \includegraphics[width=0.3\textwidth]{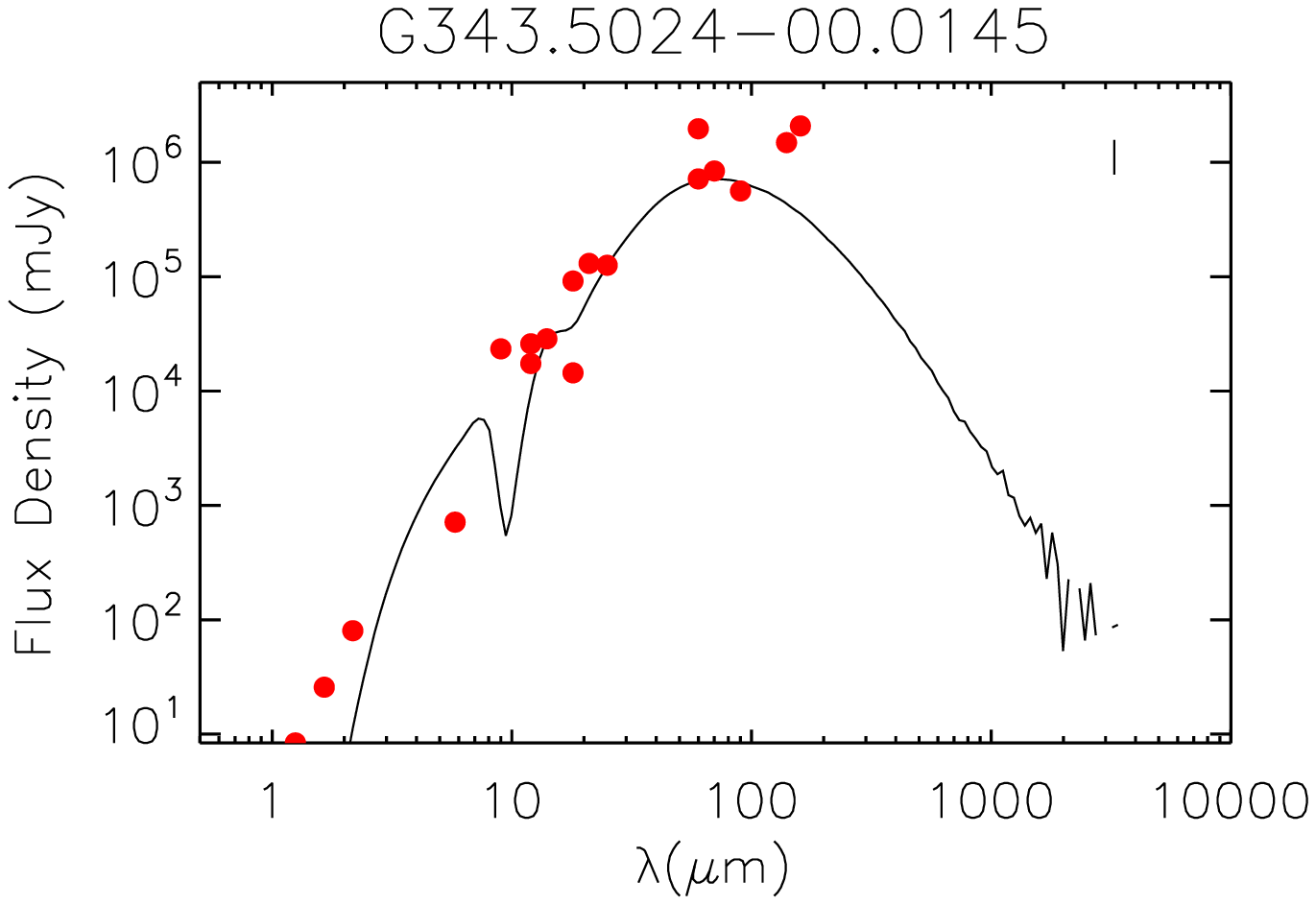}\\
        \includegraphics[width=0.3\textwidth]{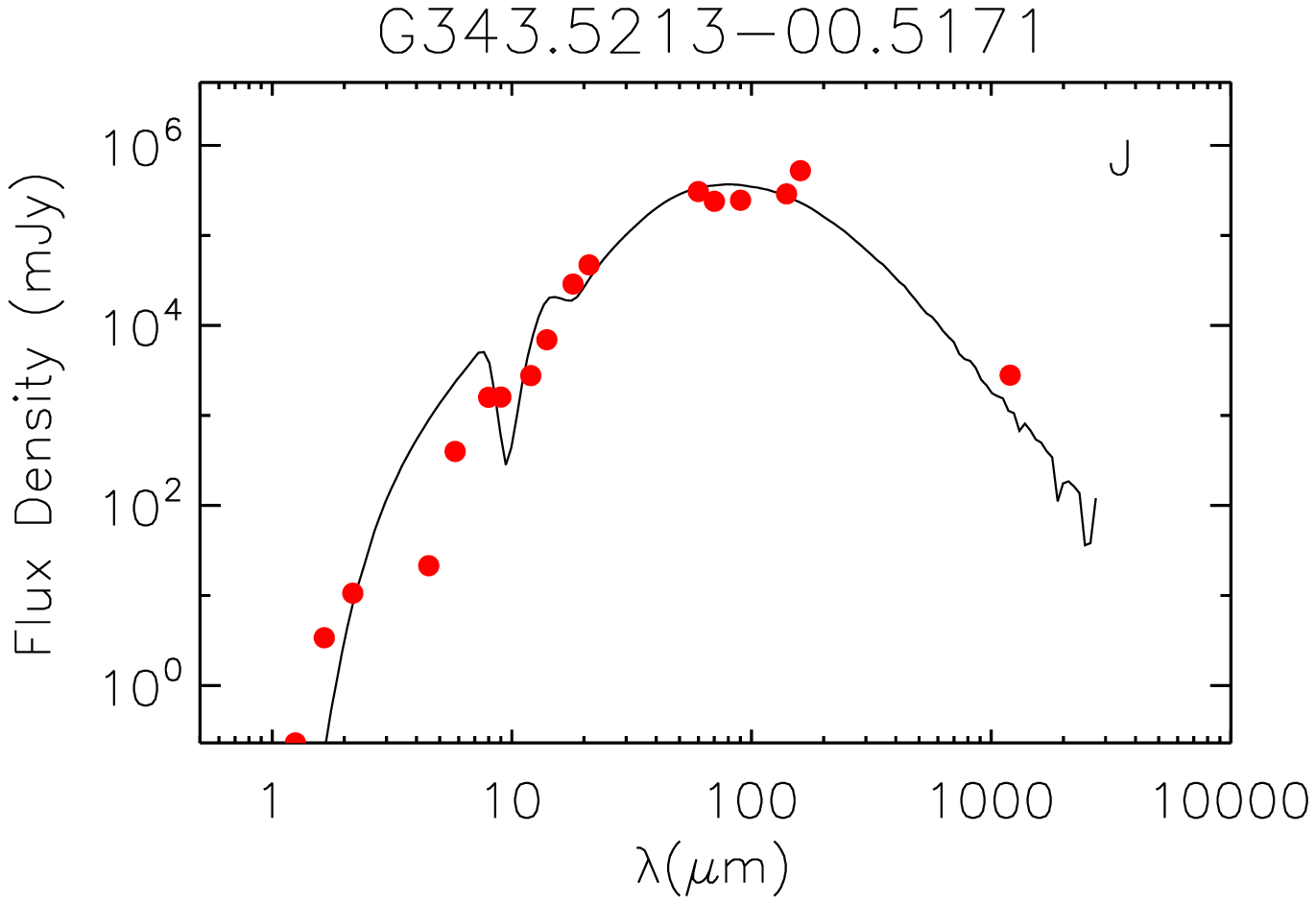}&
        \includegraphics[width=0.3\textwidth]{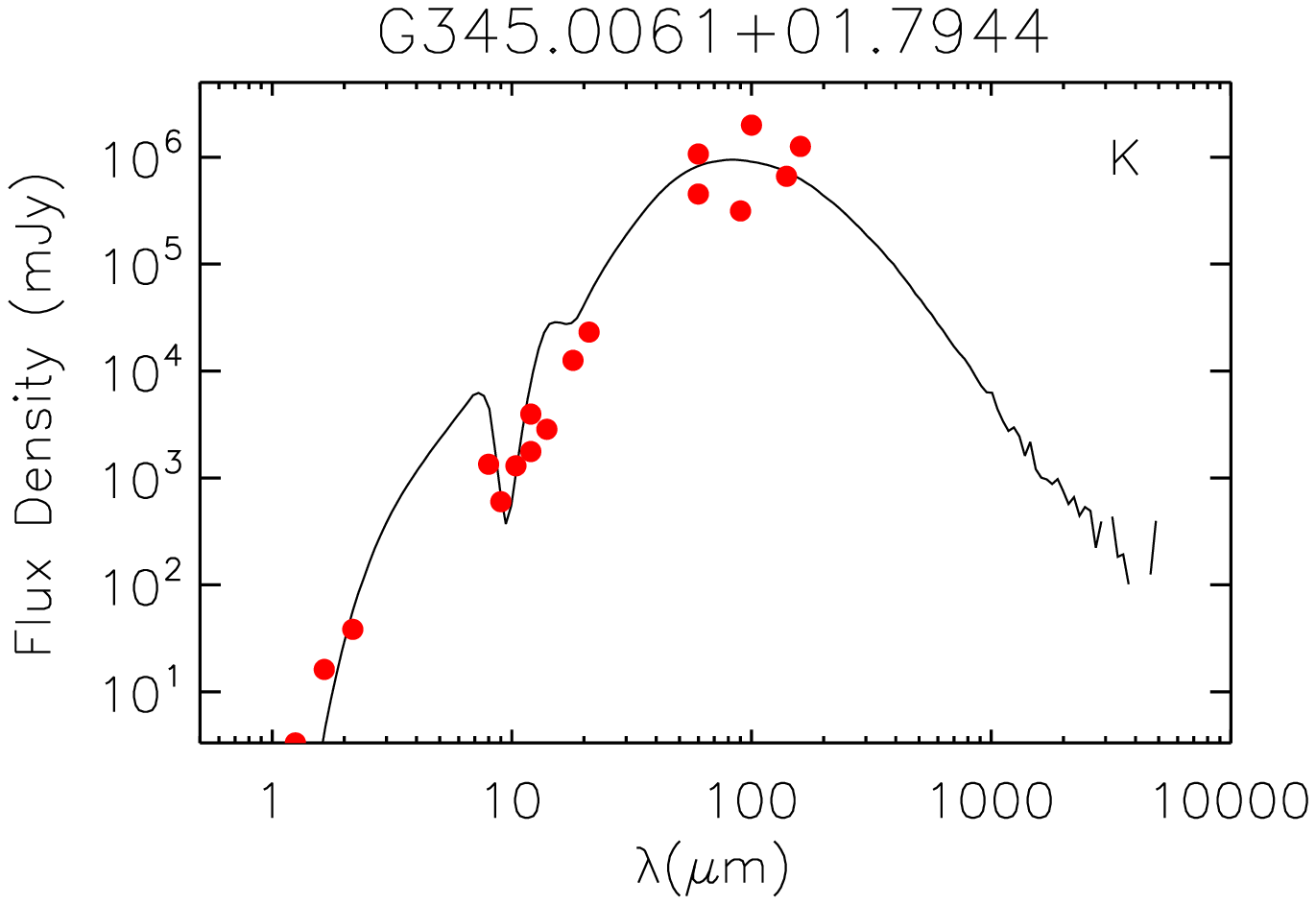}&
        \includegraphics[width=0.3\textwidth]{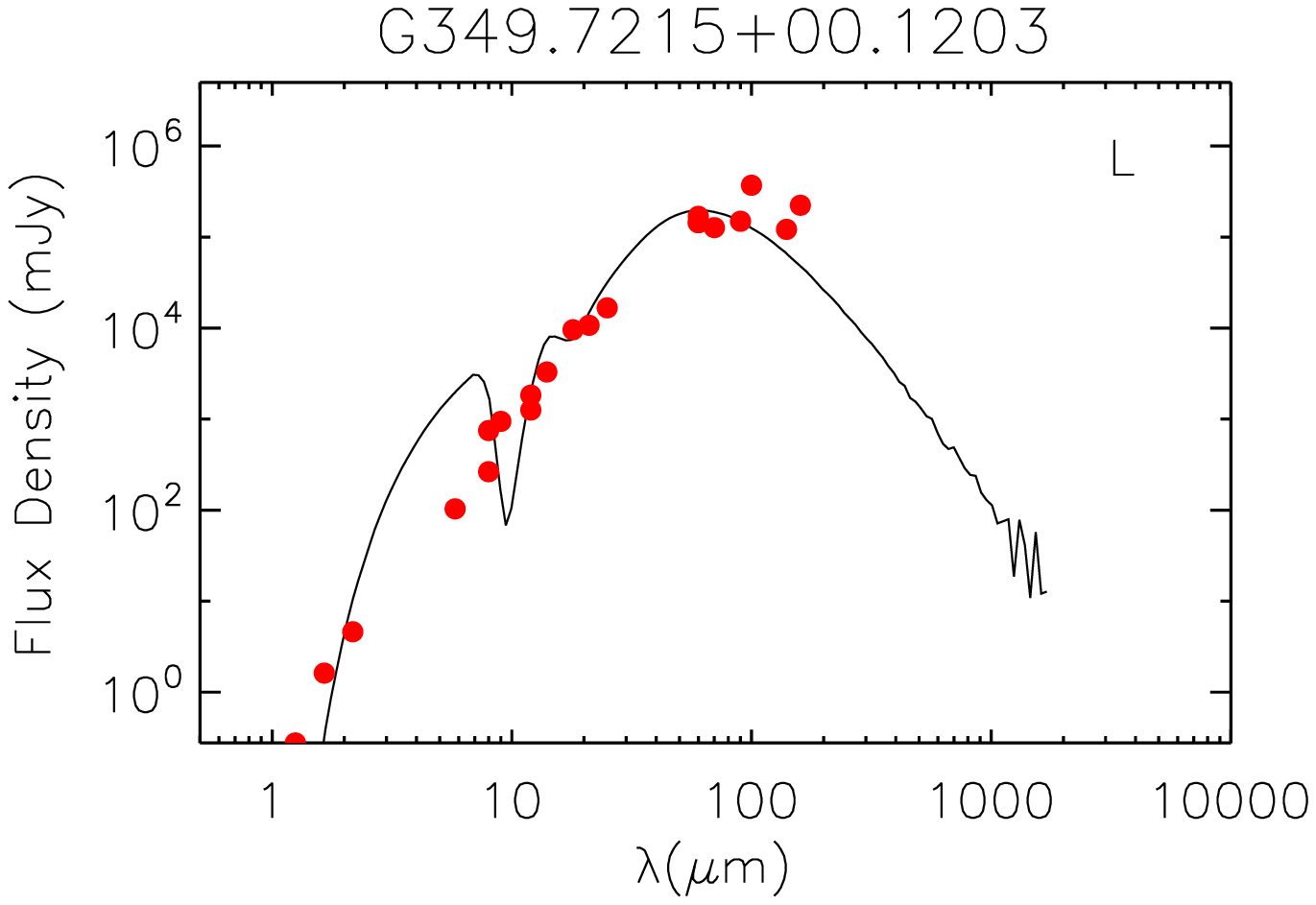} \\ 
      \end{tabular}
      \caption{The SEDs of the resolved MYSOs compared to the final model SEDs\label{seds}.}

    \end{center}
  \end{figure*}
\end{center}

\begin{center}
  \begin{figure*}[!h]
    \begin{center}
      \begin{tabular}{l l l}
        \includegraphics[width=0.3\textwidth]{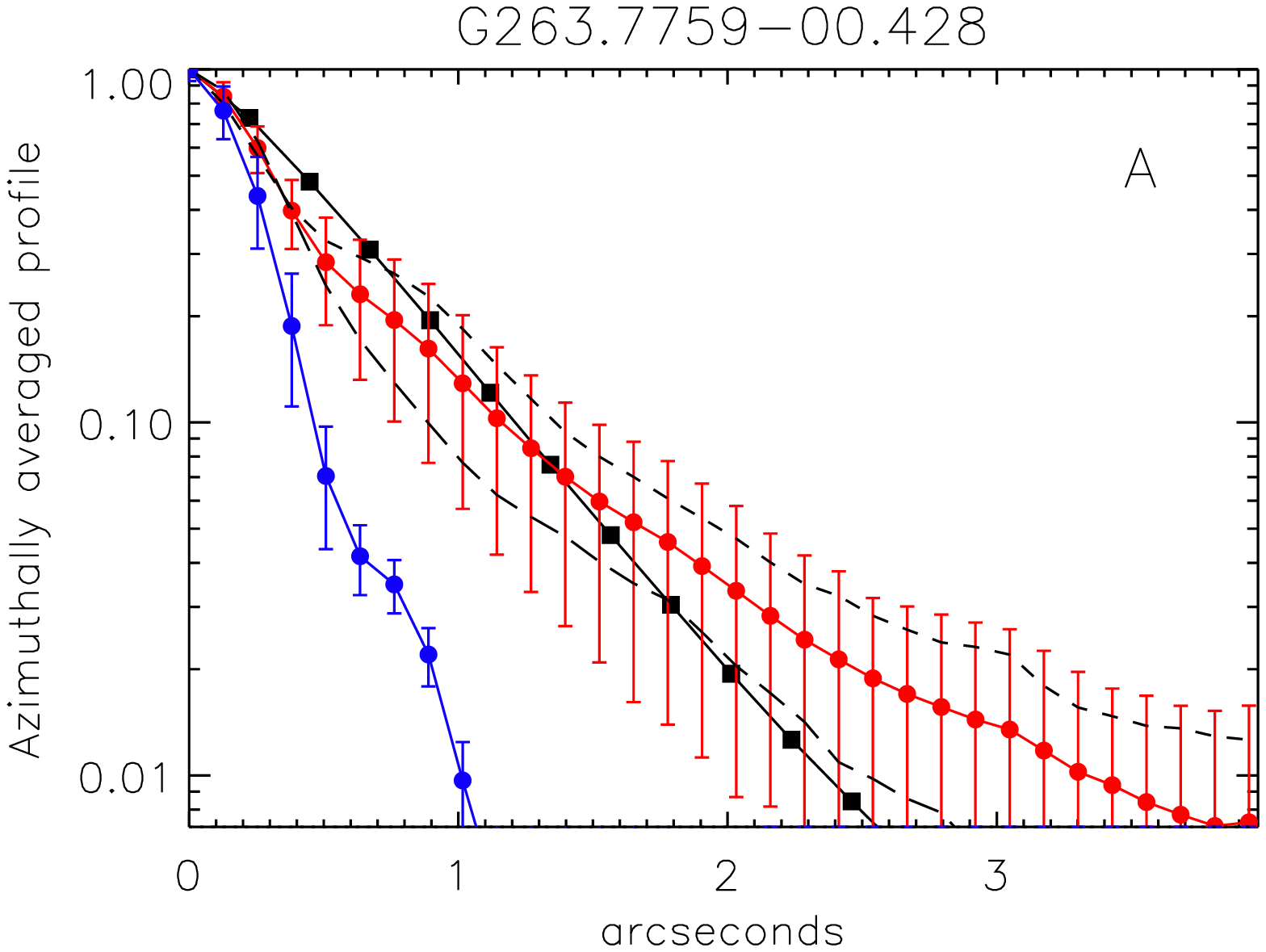} &
        \includegraphics[width=0.3\textwidth]{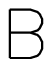} &
        \includegraphics[width=0.3\textwidth]{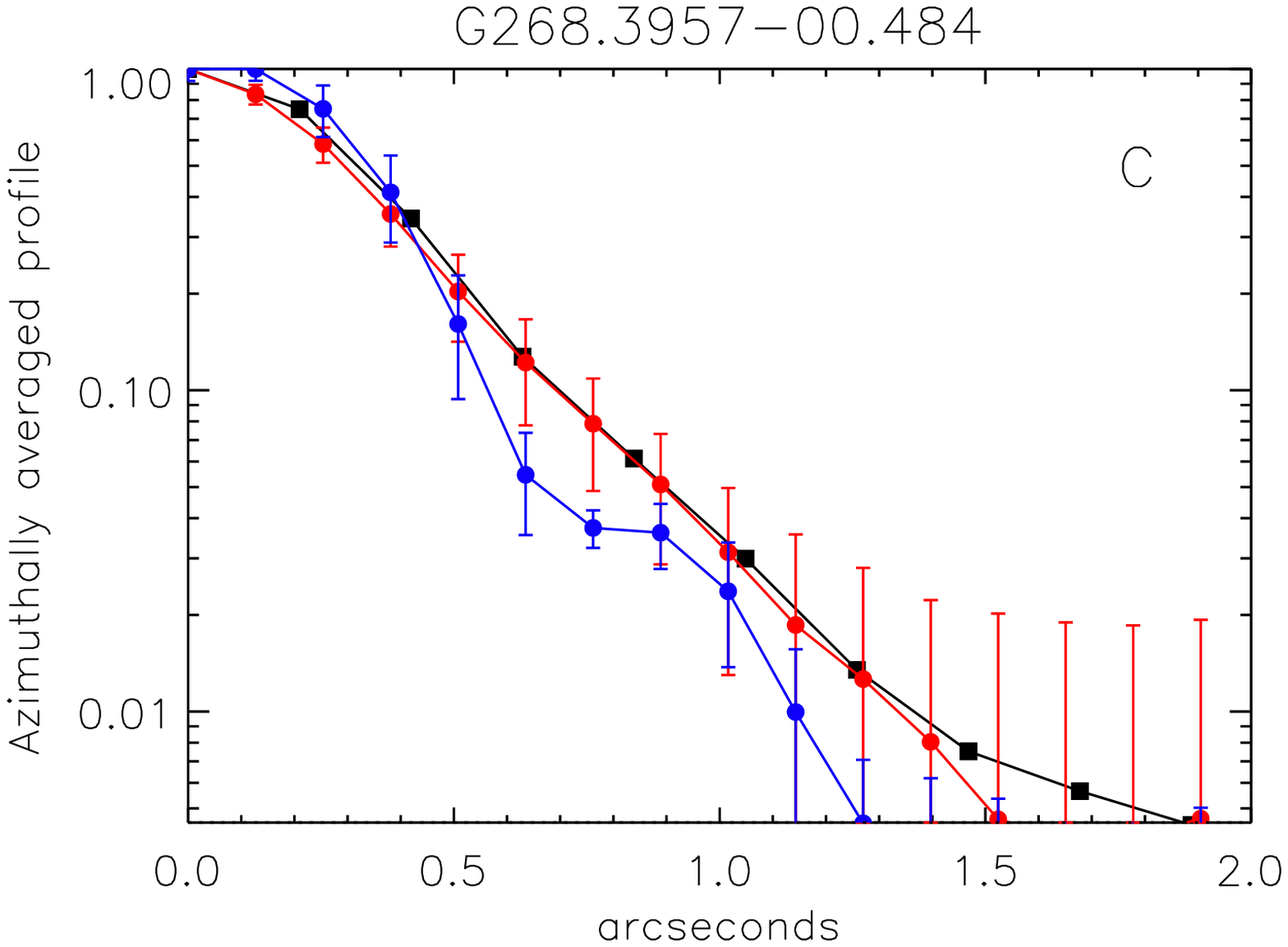} \\
        \includegraphics[width=0.3\textwidth]{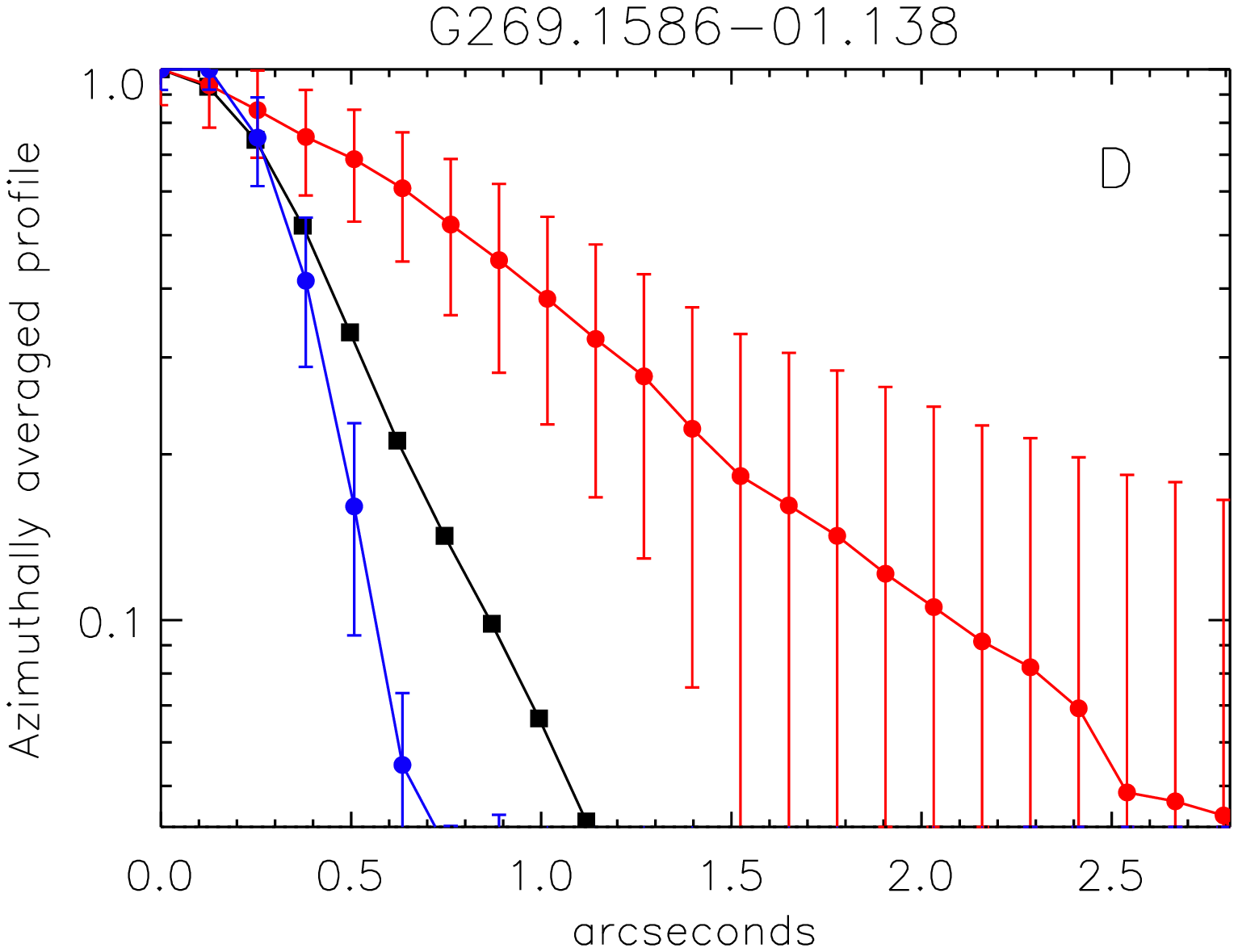}&
        \includegraphics[width=0.3\textwidth]{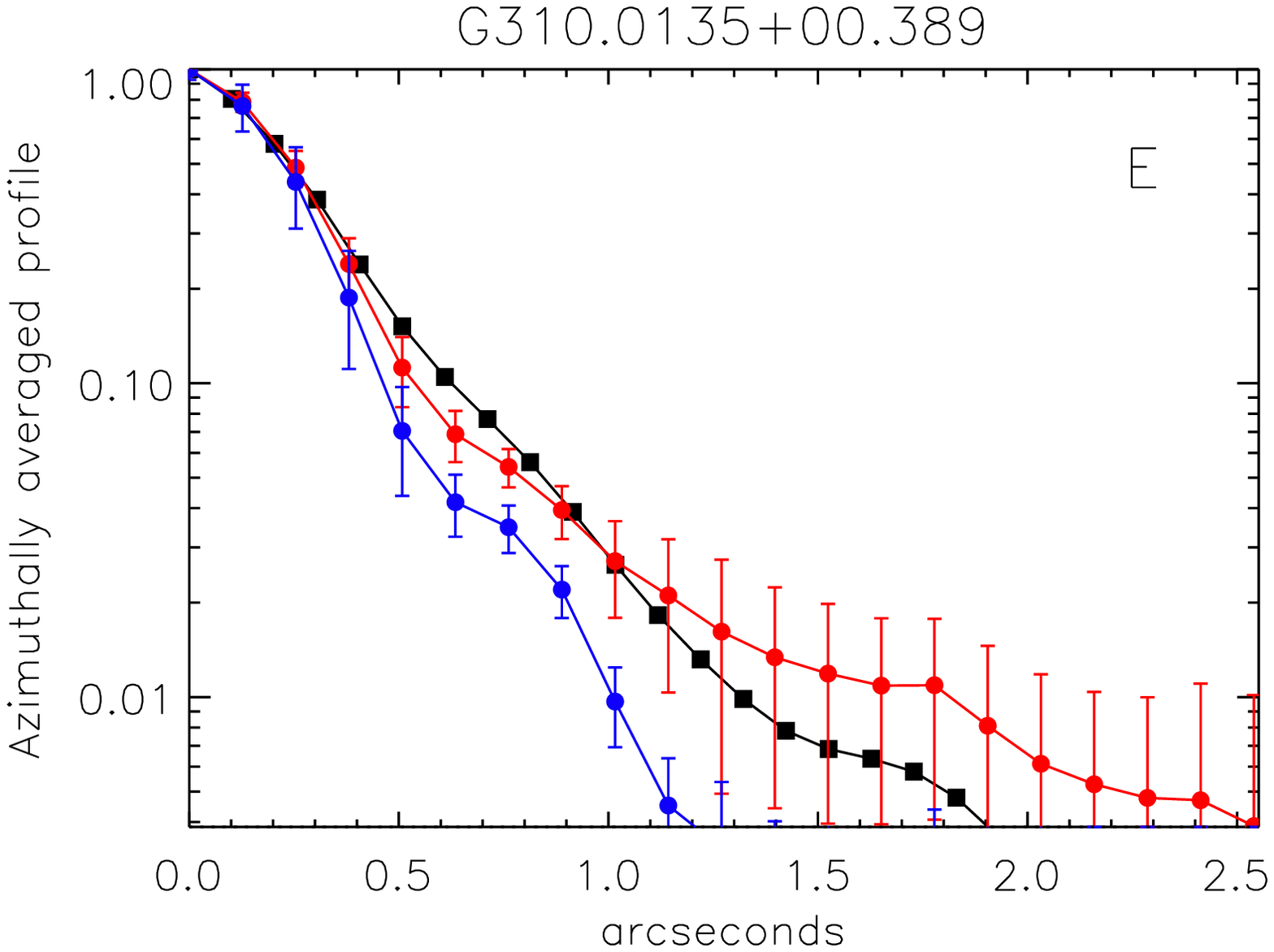}&
        \includegraphics[width=0.3\textwidth]{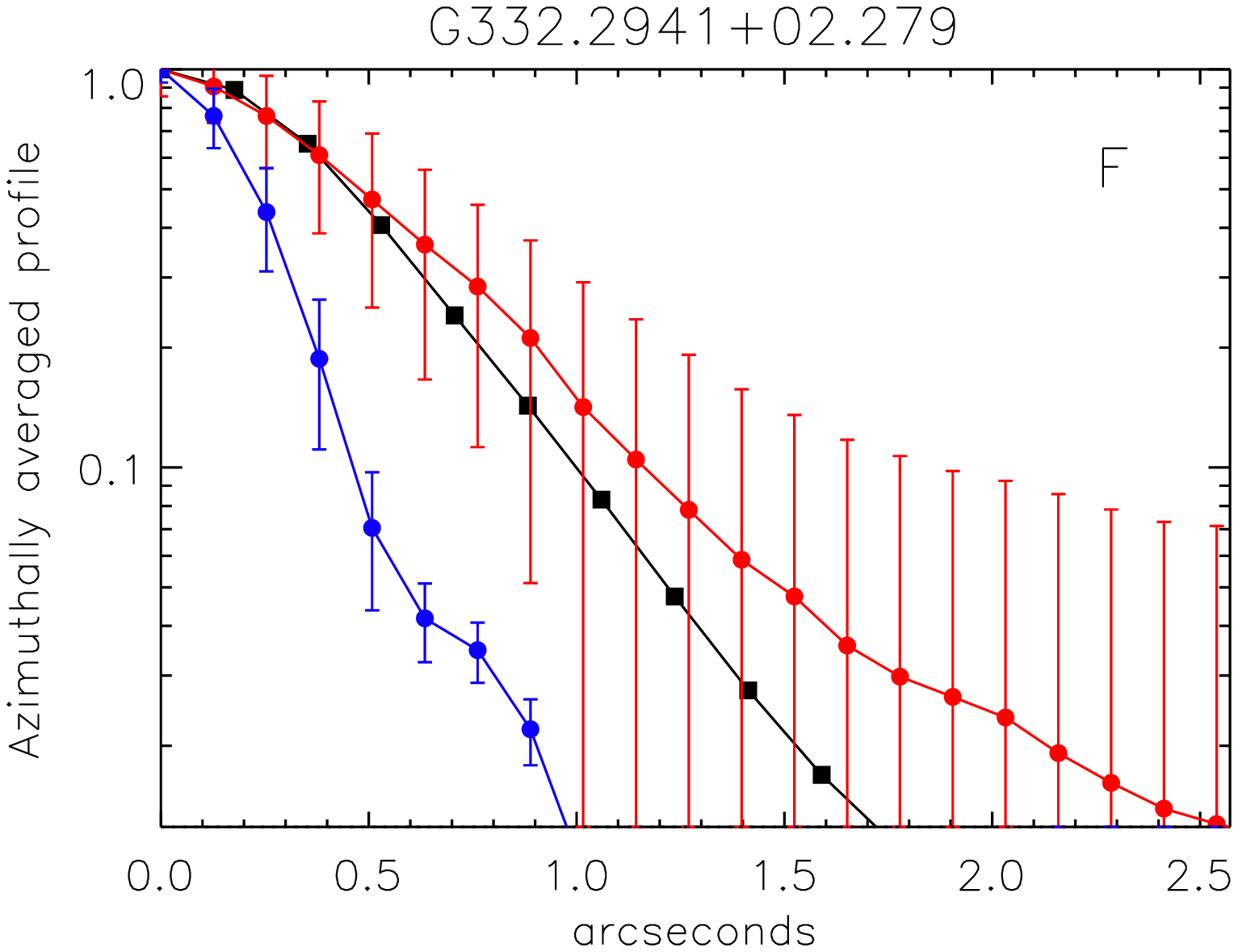}\\
        \includegraphics[width=0.3\textwidth]{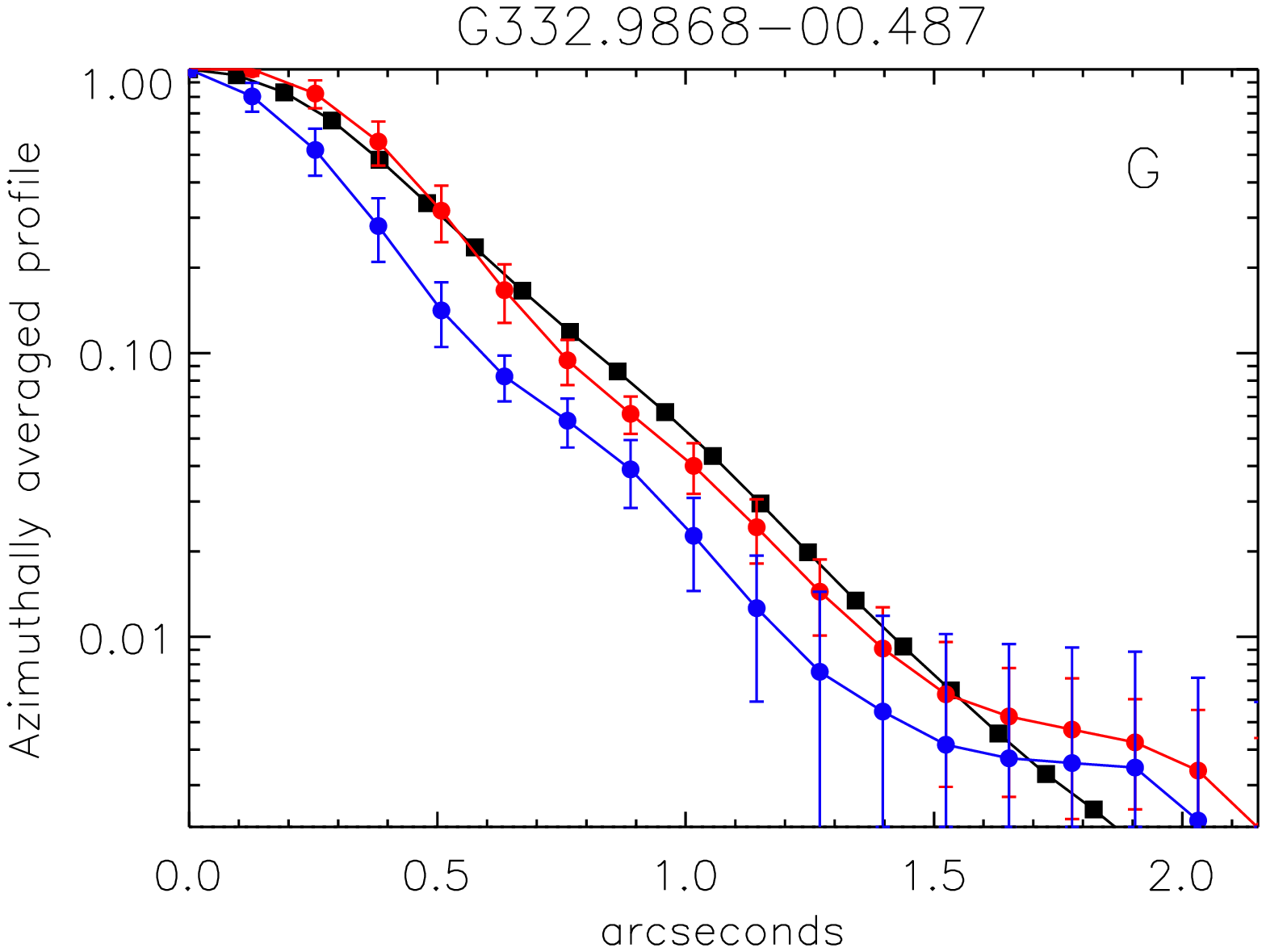}&
        \includegraphics[width=0.3\textwidth]{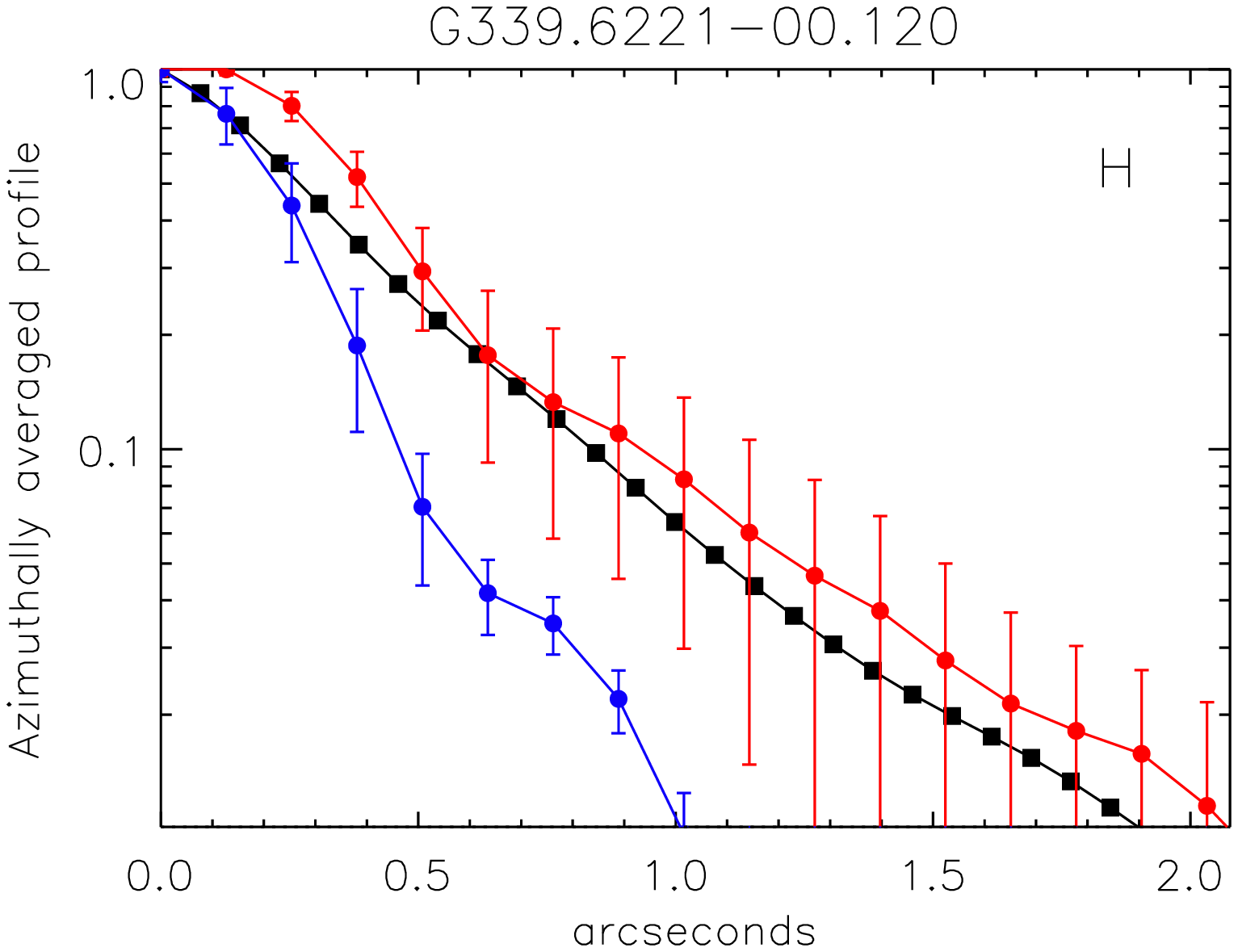}&
        \includegraphics[width=0.3\textwidth]{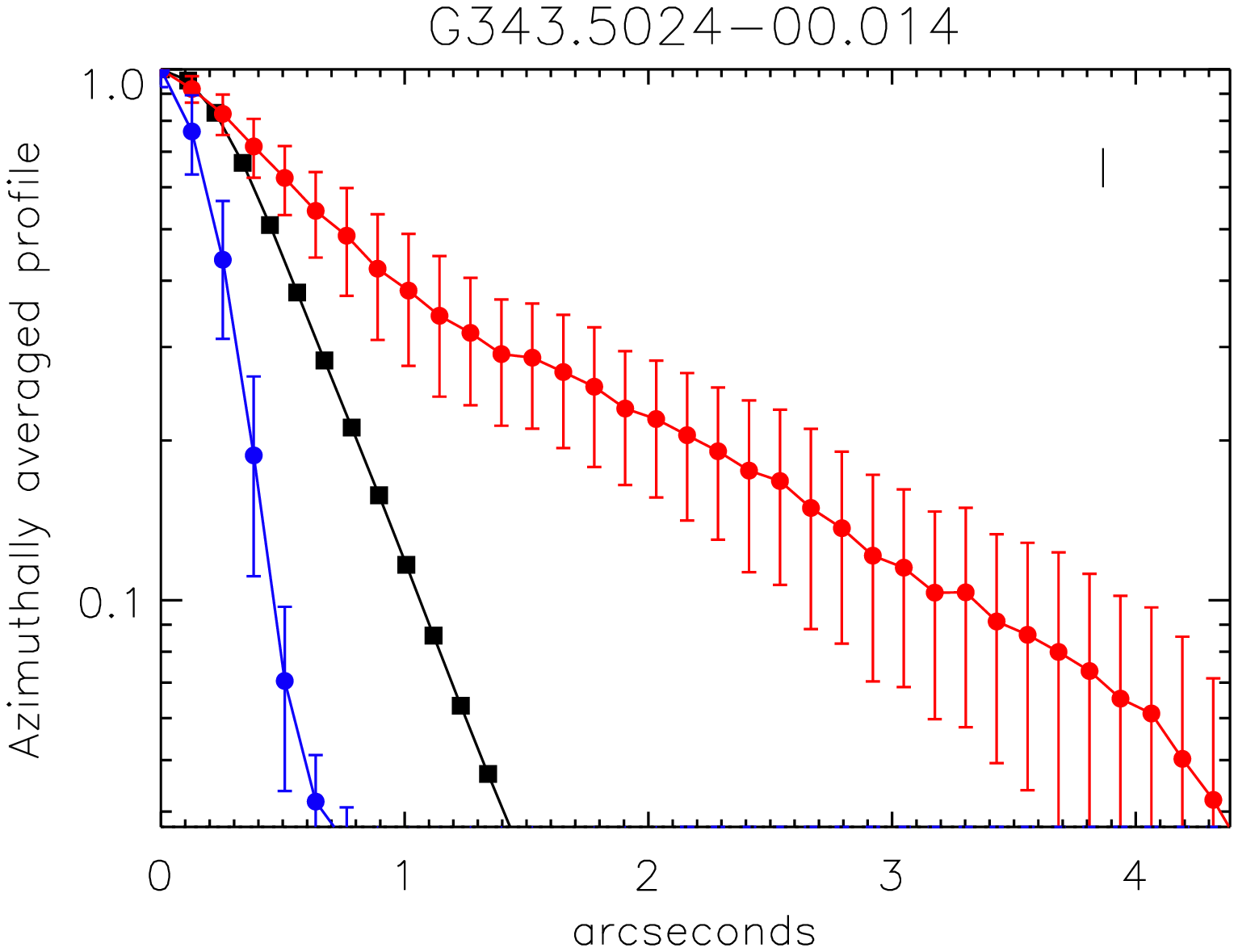}\\
        \includegraphics[width=0.3\textwidth]{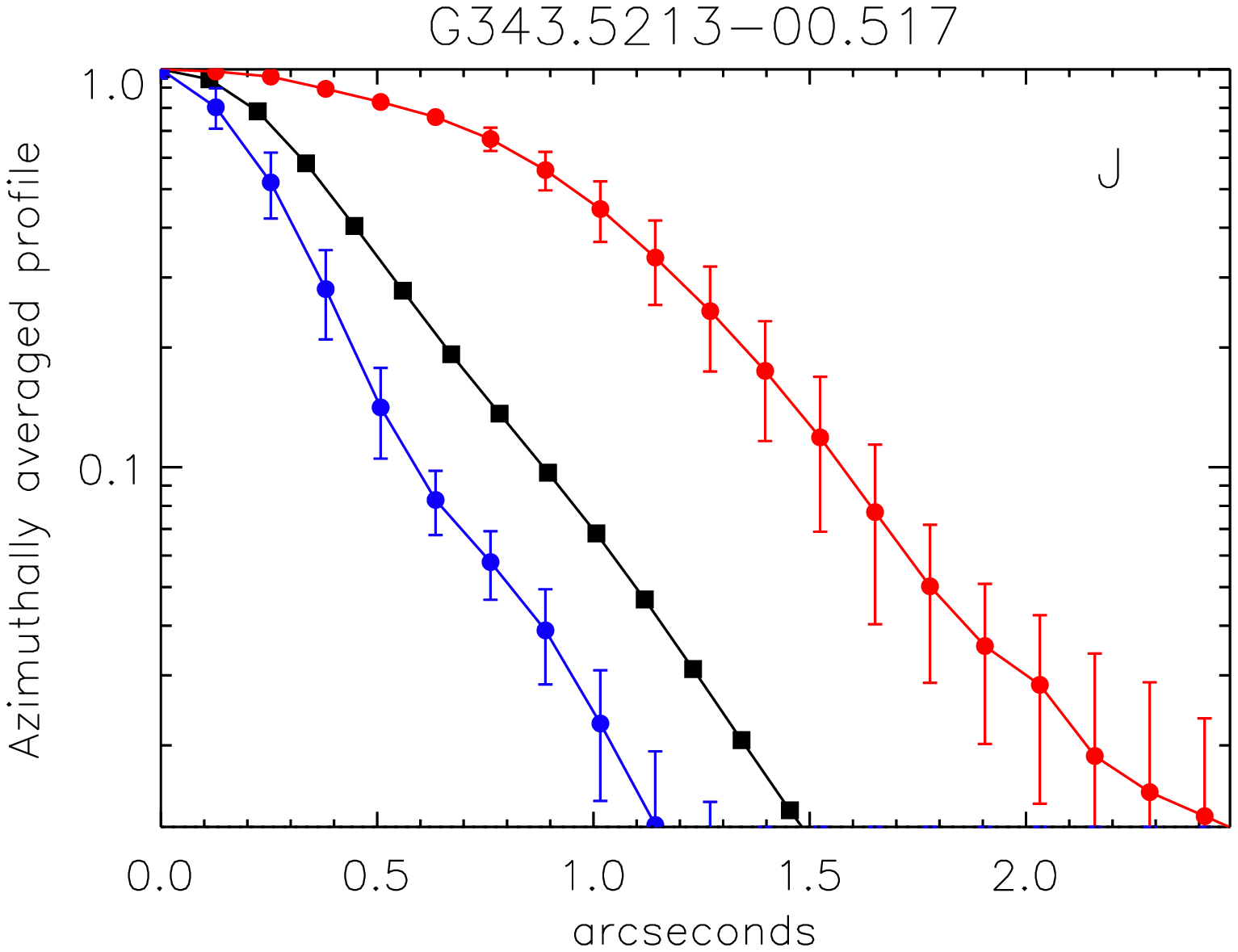}&
        \includegraphics[width=0.3\textwidth]{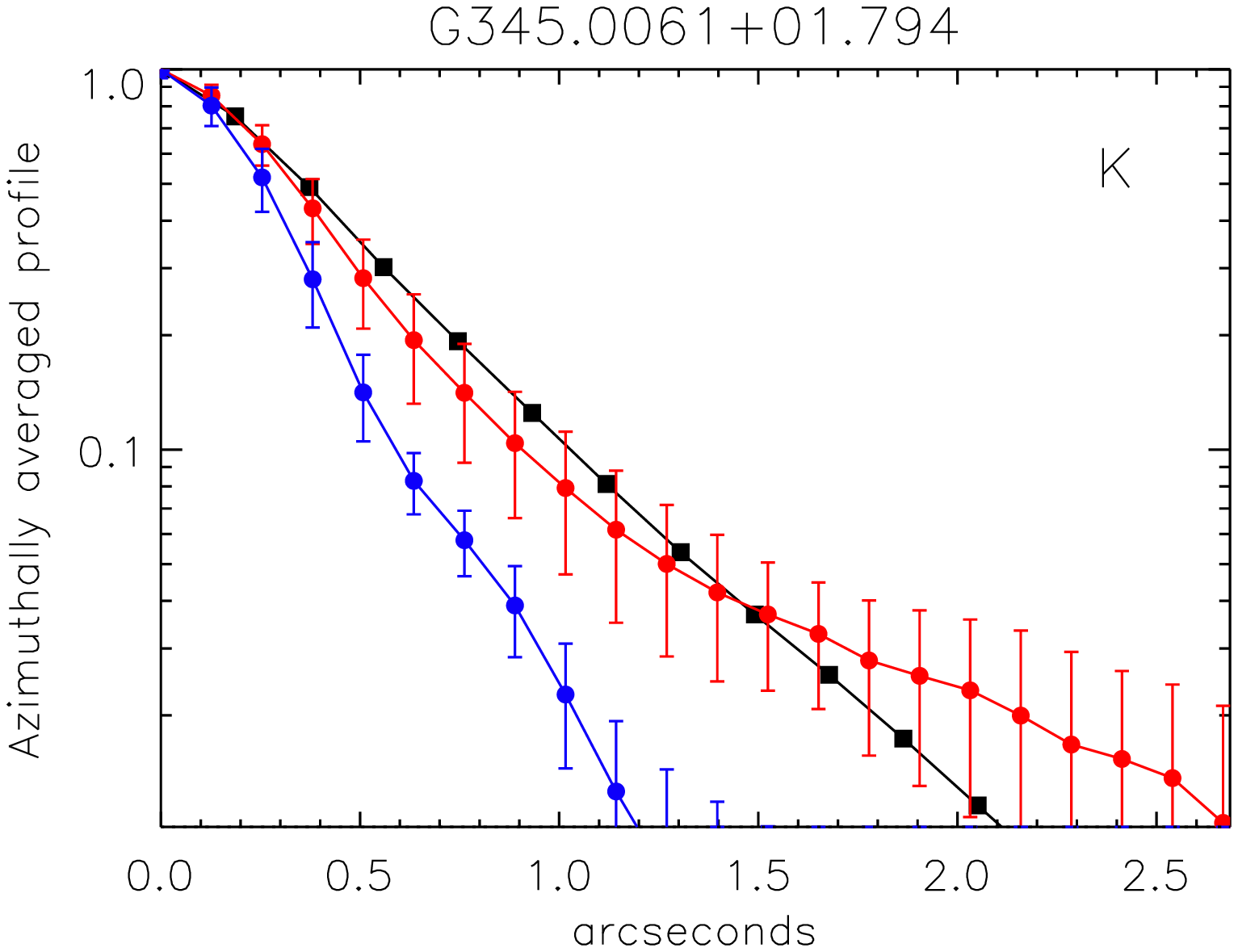}&
        \includegraphics[width=0.3\textwidth]{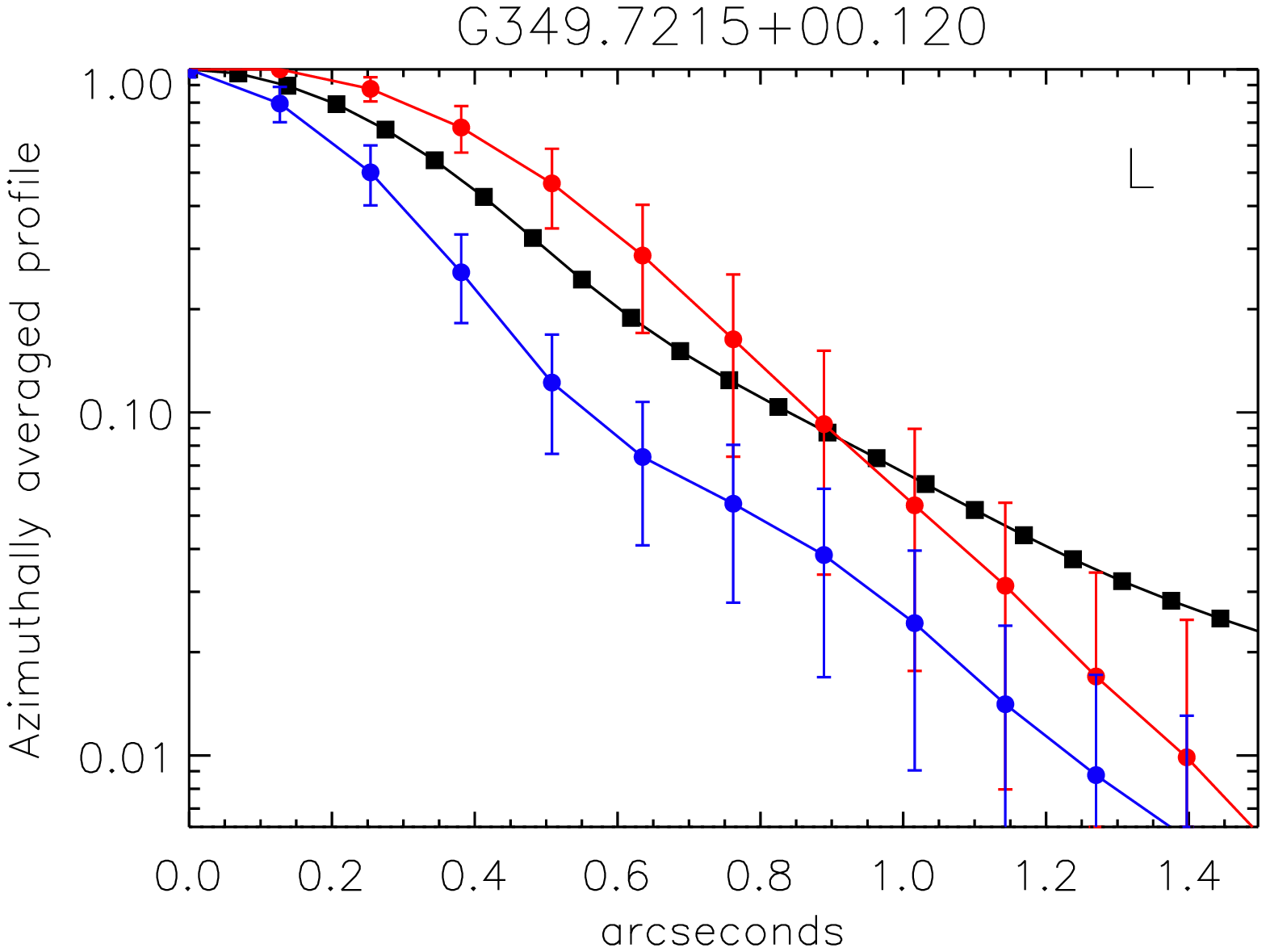} \\ 
      \end{tabular}
      \caption{The azimuthally averaged intensity profiles of the
        resolved MYSOs (upper circles) alongside the model radial
        profiles (squares) and the associated PSF standard (lower
        circles). The lower limit to the $y$ axis is set to the
        root-mean-square noise in the background of the MYSO
        images. The upper limit of the $x$ is axis is set to the
        distance at which the MYSO profiles fall to the level of the
        background noise. The radial profile of G263.7759-00.4281 is
        also shown averaged over the upper and lower half of its
        bipolar morphology separately (short and long dashed lines
        respectively). The error bars represent the rms within a given
        annuli and thus represent an upper limit on the uncertainty in
        the flux distribution.\label{rad_profs}}
    \end{center}
  \end{figure*}
\end{center}

The final SEDs are presented in
Fig. \ref{seds} and the associated azimuthally averaged intensity
profiles are displayed in Fig. \ref{rad_profs}. The parameters of the
selected models are presented in Table \ref{pars}. Since the sample
exhibits a varied morphology (see Fig. \ref{res_im}), we discuss the
modelling results for each object individually in Appendix
\ref{notes}. We discuss the general results in the following section.

\begin{table}
  \begin{center}
    \caption{The key parameters of the individual models.\label{pars}}
    \begin{tabular}{l  c c c p{0.5cm} p{0.5cm} }
      \hline
      Object & $\frac{L_{\mathrm{Model}}}{L_{\mathrm{RMS}}}$ & $i$ & $\dot{m}$ &  \multicolumn{1}{c}{$\theta_{\mathrm{op. ang.}}$}  &  \multicolumn{1}{c}{$M_{\rm{Env.}}$}\\
      &                                          &$^{\circ}$  & $M_{\odot}{\mathrm{yr^{-1}}}$ & \multicolumn{1}{c}{$^{\circ}$} & $10^3M_{\odot}$\\
      \hline
      \hline

      \multicolumn{4}{l}{{\bf{SED and intensity profile}}}\\

      G263.7759$-$00.4281 & 1.0 & 87 & $\mathrm{1.5\times 10^{-4}}$  &  \multicolumn{1}{c}{25} & 15.5\\
      G265.1438+01.4548   & 1.2 & 32 & $\mathrm{2.0\times 10^{-4}}$  &  \multicolumn{1}{c}{25} & 15.5\\
      G268.3957$-$00.4842 & 0.5 & 30 & $\mathrm{3.5\times 10^{-4}}$  &  \multicolumn{1}{c}{10} & 9.4\\
      G310.0135+00.3892   & 1.0 & 32 & $\mathrm{7.5\times 10^{-4}}$  &  \multicolumn{1}{c}{10} & 9.8\\
      G332.2941+02.2799   & 1.4 & 87 & $\mathrm{12.5\times 10^{-5}}$ &  \multicolumn{1}{c}{20} & 12.8\\
      G332.9868$-$00.4871 & 1.0 & 15 & $\mathrm{1.0\times 10^{-4}}$  &  \multicolumn{1}{c}{10} & 9.3\\
      G339.6221$-$00.1209 & 1.2 & 60 & $\mathrm{12.0\times 10^{-4}}$ &  \multicolumn{1}{c}{12} & 11.4\\
      G345.0061+01.7944   & 1.0 & 63 & $\mathrm{4.0\times 10^{-4}}$  &  \multicolumn{1}{c}{10} & 9.4\\
      G349.7215+00.1203   & 1.4 & 65 & $\mathrm{7.5\times 10^{-4}}$  &  \multicolumn{1}{c}{15} & 11.3\\

      \multicolumn{4}{l}{{\bf{SED only}}}\\
      G269.1586$-$01.1383 & 0.8 & 60 & $\mathrm{7.5\times 10^{-4}}$  &  \multicolumn{1}{c}{45} & 32.2\\
      G343.5024$-$00.0145 & 1.3 & 85 & $\mathrm{3.0\times 10^{-4}}$  &  \multicolumn{1}{c}{10} & 9.3 \\
      G343.5213$-$00.5171 & 1.0 & 57 & $\mathrm{2.5\times 10^{-4}}$  &  \multicolumn{1}{c}{10} & 9.3\\

      \hline
    \end{tabular}
    \tablefoot{We note that the output luminosity is affected by
      inclination angle, and is thus not necessarily identical to the
      source luminosity. Parameters not listed were not varied. In the
      case of several objects, particularly those associated with
      H{\sc{ii}} regions, e.g. G343.5024$-$00.0145, the model only
      reproduces the objects' SED. {{$\theta_{\mathrm{op. ang.}}$
          denotes half the full opening angle}}.}
  \end{center}
\end{table}

\section{Modelling results}

\label{mod_res}

In general, the observed SEDs could be reproduced relatively
well. There are several cases where the observed NIR fluxes could not
be reproduced exactly. However, we do not weight the NIR fluxes
heavily as these can be strongly dependent on dust opacities and the
circumstellar geometry on small scales. In the majority of cases,
barring objects associated with H{\sc{ii}} regions, the spatial
distribution of the 20\,$\mu$m emission of the MYSOs observed can also
be reproduced by our 2D axis-symmetric radiative transfer (RT)
model. The individual model fitting results for the SEDs and the
20\,$\mu$m intensity profiles are shown in Figs.\,\ref{seds} and
\ref{rad_profs}. Clearly, because of the unequal SED coverage from
source to source, the model fidelity differs. We do not attempt to
provide a unique model for each single source but evaluate whether the
W33A model is capable of providing satisfactory fits, by changing a
few of the model's most basic parameters. From this starting point, we
find that in general, if the image exhibits a single source with an
approximately symmetric morphology, both the object's SED and the
20\,$\mu$m intensity profile are well reproduced. By extension of the
case for W33A, the majority of the 20\,$\mu$m emission in these
instances is very likely emission from warm dust in cavity walls.

\smallskip

When an object's morphology appears either very complex or cometary, the
model generally fails to reproduce the intensity profile. For example,
this is the case for G343.5024$-$00.0145, which is a known
ultracompact H{\sc{ii}} region. For this object, a decent fit to the
SED (panels I in Figs. \,\ref{seds} and \ref{rad_profs}) predicts a
``MYSO'' structure that is much more compact than the one observed,
even with extreme values for the inclination and opening angle. A
cometary appearance alone will often indicate an ionized nature for
the region, and this is generally confirmed via radio continuum
observations. That the model cannot reproduce the observations of
MYSOs associated with H{\sc{ii}} regions is perhaps to be expected as
H{\sc{ii}} regions represent a source of flux not considered by the
model. Furthermore, the appearance of a H{\sc{ii}} region likely
signifies the end of the embedded accretion phase of a MYSO, and it is
this phase the model is designed to represent.

\smallskip

We give an overview of the values for the varied parameters in
Table\,\ref{pars}. The opening angles of the best-fitting models are
generally narrow ($\sim10 \degr$). This implies that the outflows of
the MYSOs observed are relatively well collimated. The envelope infall
rates of the models, a parametrisation of the density distribution,
are generally of the order
$\mathrm{10^{-4}}M_{\odot}\mathrm{yr^{-1}}$. The corresponding values
of visual extinction, $A_V$, are typically between 10 and 30, in
agreement with the measured $A_V$s of MYSOs \citep{Porter1998},
although several of the more inclined models have $A_V$s in excess of
100. The total masses of the envelopes are generally of the order
10,000~$M_{\odot}$. {\color{black}We note that these are dependent on
  the envelope outer radii, which are set at $\mathrm{5 \times
    10^5}$~AU. In the case of the W33A model, reducing the outer
  radius by a factor of 2 decreases the total mass by a factor of
  approximately 5. This is a result of adopting a low but constant
  density, $\rho = 8 \times 10 ^{-20} \mathrm{g\,cm^{-3}}$, within the
  outflow cavity.} The envelope masses calculated are comparable to
the clouds that the objects are located in \citep[see][and references
therein]{Urquhart2011}. Even allowing for an uncertainty of a factor
of a few, this indicates that these objects are still heavily embedded
in their natal material and their environment still contains a large
reservoir of material available to form further stars.

\smallskip

\section{Discussion}
\label{disc}
\subsection{On the mid-IR emission of MYSOs}

Over the past decade, observational evidence indicating that the
formation of massive stars, up to a mass of at least $\sim
30\,M_{\odot}$, is accompanied by phenomena characteristic
of low mass SF has accumulated. Such phenomena include:
circumstellar discs \citep[see
e.g.][]{Kraus2010,Davies2010,Masque2011,Goddi2011}, molecular outflows
\citep[see e.g.][]{Beuther2002,Cyganowski2009} and jet activity
\citep[see e.g.][]{Rodriguez1994,Guzman2010}. 

\smallskip

Despite the many common characteristics of low and high mass SF, some
observational phenomena seem to be exclusively associated with young,
embedded high-mass stars. One example is the Class II 6.7-GHz methanol
maser, whose activity is uniquely associated with massive star forming
regions \citep[e.g.][]{Walsh1997}. These sites evidently produce
sufficient IR radiation in order to pump this transition, unlike sites
of low mass SF. It was initially thought that this maser activity
originated in circumstellar discs. As a result, a correlation between
the extension of several MYSOs in the MIR and their maser emission led
to the suggestion that the MIR emission of MYSOs traces discs
\citep[see e.g.][]{DeBuizer2000}. However, observations with 8m class
telescopes were able to spatially resolve the MIR emission of a number
of masing sources to reveal that their MIR emission is aligned with,
rather than perpendicular to, their large CO molecular outflows
\citep[see e.g.][]{DeBuizer2006,DeBuizer2007}. These findings
associate the MIR emission of massive YSOs with outflow cavities,
rather than circumstellar discs. Nevertheless, the kinematic diversity
of methanol masers in a large sample of massive outflow sources does
not support a uniquely identifiable source within the close MYSO
environment \citep[][]{Cyganowski2009,Moscadelli2011}. Therefore, the
source of the MIR emission of MYSOs has yet to be conclusively
established.

\smallskip

Recently, several massive young stellar objects with compact MIR
emission were investigated using long-baseline interferometry. In
several cases, the spatially resolved $N$-band emission on scales of
100~AU was found to be consistent with thermal envelope emission (de
Wit et al. 2007; Linz et al. 2009\nocite{Linz2009}) and in particular
with emission arising from the cavity walls (de Wit et al. 2010).
However, in some other cases, the mas-scale MIR emission of MYSOs is
suggested to be (at least partially) located in a circumstellar disc
\citep[see e.g.][]{Follert2010,deWit2011,Grellmann2011}. In other
words, there are hints that the $N$-band emission, resolved with
interferometry, is not completely dominated by envelope emission. The
analysis of the $Q$-band emission presented in this paper favours the
cavity wall emission interpretation for a sample of MYSOs. Upon first
inspection, this appears at odds with the scenario of a disc
dominating the $N-$band emission. Here, we consider the possible
contribution to the $Q-$band emission by a circumstellar disc in order
to assess whether our finding agrees with theoretical expectations.

\smallskip

The recent study by \citet{Zhang2011} discusses in detail the
appearance of embedded young massive stars in the IR wavelength
region. The authors use a dust radiative transfer code, as we
do. {{However, their geometry differs from the one employed in this
    paper in that their code includes an accretion disc that extends
    from the stellar surface to the centrifugal radius}}. Therefore,
their results allow us to assess the possible disc contribution to the
$Q-$band emission of MYSOs. The RT calculations of \citet{Zhang2011}
reveal that outflow cavities have a large influence on the shape of
the SED. Furthermore, in their calculation for a $60^{o}$ inclination
(see e.g. their Fig. 9), the outflow cavities are clearly significant
features in the images, up to a wavelength of 70\,$\mu$m. In the NIR,
scattered light from the cavities makes a significant contribution to
the near-IR emission. In the $N$-band, thermal emission from the disc
and cental object are also evident. In the $Q$-band however, the
emission from the base of the cavity clearly dominates the flux.

\smallskip

The relative contributions of the disc and envelope to the modelled
SED depend critically on exactly those parameters with which we fit
our observations, i.e.  opening angle, mass infall rate and
inclination. We note that the opening angle applied by
\citet{Zhang2011} is rather wide ($2\alpha \approx 90^{o}$), which
would imply an object in a rather advanced phase of formation,
following the expected widening of the opening angle as function of
time \citep[see e.g.][]{Kuiper2010}. Smaller opening angles increase
the extinction towards the central regions and thus favour the
dominance of cavity wall emission in the total MIR emission. A similar
reasoning applies for the mass infall rate. It is therefore not
surprising that the observed MYSOs, which are believed to be in a less
evolved phase than the phase represented by the fiducial model of
\citet{Zhang2011}, are dominated by the cavity wall emission. This
argues against a large effect, if any, by a circumstellar disc on our
VISIR images. This conclusion can also be inferred from the spectral
energy distribution of an accretion disc, since it peaks at
wavelengths much shorter than 20\,$\mu$m. Furthermore, the envelope
contains much more cool material than does any reasonable accretion
disc, whose sizes for MYSOs are estimated to be less than 500\,AU
\citep[see
e.g.][]{Patel2005,Hoare2006,Reid2007,Izas2007,deWit2011,Goddi_2011}. We
conclude that it is reasonable to expect that dust emission from the
envelope (the cavity walls) dominates the continuum SED at
wavelengths longward of the $N$-band. This is in agreement with our
finding that the observed extension of the $Q-$band flux can be
modelled as emission from outflow cavity walls.

\subsection{The structure of MYSO envelopes: rotation versus outflows}

We have shown that the morphology of the 20~$\mu$m emission of MYSOs
is consistent with that predicted by models incorporating outflow
cavities. We followed a similar methodology to DW09 and have improved
on the modelling part of their analysis. DW09 compared their images to
1-D radiative transfer models. This was motivated by the rather
spherical appearance of most sources, which suggests that their MIR
emission originates from a relatively symmetrical dusty envelope. They
found that, in general, their data were best fit by an envelope with a
density gradient of $\rho \propto r^{-1}$. This density law is
shallower than that obtained via a similar analysis performed in the
sub-mm \citep[see e.g.][]{Mueller2002}. It was therefore suggested
that this flattening of the density law towards smaller spatial scales
could be evidence for the onset of rotational support of the envelope
at approximately 1000~AU. This suggestion is probably invalidated by
the results presented in this paper based on more realistic RT models.

\smallskip

Thermal emission from an envelope with evacuated outflow cavities fits
the SED and the $N$-band (de Wit et al. 2010) and $Q$-band (this
paper) morphology of the luminous MYSO W33A. Additionally, in this
particular case, the envelope model also fits the observed
morphologies of the near-IR scattering nebula and the 350\,$\mu$m
emission. This provides strong evidence in favour of this model and
our explanation of the 20~$\mu$m emission (warm dust in the cavity
walls). Moreover, the W33A image (Fig.\,\ref{test_image}) is slightly
extended towards the SE which is the direction of the blue lobe of the
large scale outflow, reinforcing this interpretation. Importantly, we
note that W33A is not the only case in our sample for which the
extension at 20~$\mu$m is in the direction of the outflow
position angle (see for example the sources G263.7759$-$00.4281 and
G332.2941+02.2799). Therefore, our association of the observed MIR
flux with outflow cavity walls can be substantiated.

\smallskip

The brightness of the cavity walls, as exemplified in panel a) of
Fig.\,\ref{test_image}, is a consequence of the density contrast
between the in-falling envelope and the outflow cavities. The
radiation of the central source cannot penetrate far in the dense
envelope, but can heat dust in the walls of the outflow cavity to
large radii. Once the model images are convolved with an instrumental
PSF, it is no longer obvious that the flux traces outflows, despite
the source being resolved. The flux from the outflow cavity walls can
appear more extended than symmetric envelope emission. A centrally
concentrated density distribution (steep density law) in a spherical
envelope produces a small thermal emission region in the MIR. The
shallower the density law the larger the (normalised) emission region
becomes. Therefore, the influence of the outflow cavities can mimic
the effect of a shallow density gradient. We verified that the sizes
of the emission regions differ strongly between a $-2$ and $-1$
density law, but the difference between a flat density law and a $-1$
density law are relatively minor. Still the SEDs differ
significantly. The steeper density law produces higher MIR fluxes. As
an exercise, we fitted the standard 2D W33A model with a spherical
model and find that the SED and intensity profiles are matched best by
a density law between $-0$ and $-1$. This explains the results
presented in DW09. Given the more realistic nature of the 2D models we
have employed, we suggest that the preference for shallow density laws
found by DW09 is the result of much of the MIR emission of MYSO emanating from
warm dust in the cavity walls. An explanation in terms of rotational
support on scales of $\sim$1000~AU cannot be substantiated.

\subsection{SED fitting}

Fitting spectral energy distributions has long been used as a tool to
provide important constraints on the morphology of circumstellar
environments \citep[see e.g.][]{Guertler1991}. Still, the inclusion of
spatially resolved observations at different wavelength regimes is
indispensable for a correct interpretation of the observed emission
\citep[see e.g.][]{vdt2000}. This is because SED fitting can
result in degenerate solutions. To assess the added value provided by
the VISIR images presented in this paper, we use the results
presented in \citet{M2011}. These authors used the grid of models
provided by \citet{Robitaille2007} to fit the SEDs of a large
percentage of MYSOs from the RMS survey in order to determine their
bolometric luminosity. The grid of \citet{Robitaille2007} was created
using the same RT code of Whitney et al. which we use in this
paper. Therefore, we used the code to create images for the best
fitting models found in \citet{M2011} (in the filter used) and assessed
whether the model images are consistent with our observations.

\smallskip

\begin{center}
  \begin{figure}
    \begin{center}
      \includegraphics[width=0.4\textwidth]{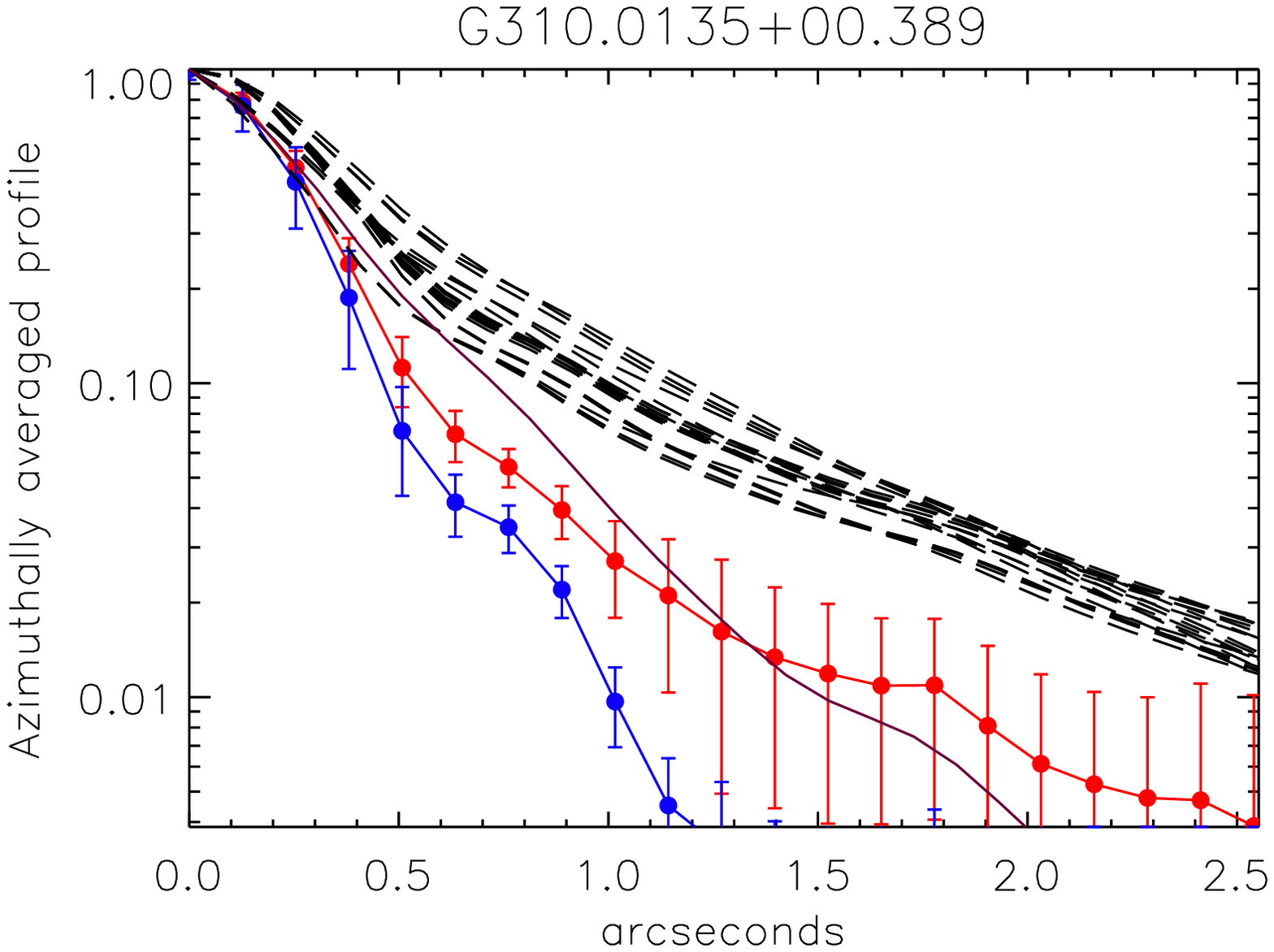}
      \caption{The azimuthally averaged intensity profile of G310.0135
        (upper circles) compared to a PSF standard (lower circles) and
        the models that fit the SED of this object from
        \citet[][dashed lines]{M2011}. The solid line is the radial
        profile of the model of \citet{Kraus2010}. The error bars
        represent the rms within a given annuli and thus represent an
        upper limit on the uncertainty in the flux
        distribution.\label{test}}
    \end{center}
  \end{figure}
\end{center}

To illustrate the results of this exercise, we focus on the object
G310.0135+00.3892. This is an object of particular interest as it is
the object studied by \citet[][]{Kraus2010} via a high angular
resolution VLTI aperture synthesis image in the $K$-band. We generate
20\,$\mu$m images of the set of models that best fit the object's SED
as found by \citet{M2011}. The result is shown in Fig. \ref{test}. The
SED fitting tool preferentially returns models with large cavity
opening angles ($\sim30-50 \degr$), but these all fail to reproduce
the observed intensity profile. The best fitting model of
\citet{Kraus2010}, which is also based on SED fitting with the model
grid of \citet{Robitaille2007} and the code of Whitney et al.,
provides a better match. However, it still over-predicts the flux in
the central 1\arcsec~from the source (see Fig. \ref{test}). This is
likely due to the large opening angle of the Kraus et al. model
($\mathrm{40^{\circ}}$) generating a MIR emission region which is too
extended. The VISIR observations clearly demonstrate that, for the
assumed intermediate inclination, a narrow opening angle is
required. We conclude that the spatially resolved VLT-VISIR images at
20~$\mu$m provide valuable constraints on the nature of the envelopes
surrounding young massive stars.

\section{Conclusions}

\label{conc}

In this paper we present diffraction limited, MIR imaging of a sample
of massive young stellar objects drawn from the RMS survey. By
comparing the spatially resolved images of the MYSOs to 2D radiative
transfer models, we constrain the structure of the envelopes that
surround them. Here we list the salient results.

\begin{itemize}

\item[--]{We spatially resolve the MIR (19.5 or 24.5~$\mu$m) emission
  of 14 MYSOs. The remainder of the MYSOs observed (6) are unresolved
  at the current resolution ($\sim$0.6\arcsec).}

\item[--]{The model of the MYSO W33A that we have developed previously
    can, in most cases, reproduce the images and SEDs of the MYSOs. This
    is not the case for objects associated with a H{\sc{ii}}
    region. That the model can reproduce the infrared emission of a
    sample of MYSOs suggests that the circumstellar environments
    of MYSOs are relatively uniform. The large
    envelope masses of the models are consistent with the
    identification of MYSOs as objects in their accretion phase.}

\item[--]{It is found that the extent of the MIR emission of the
    spatially resolved MYSOs can generally be attributed to warm dust
    located in the walls of outflow cavities. We suggest that the
    relatively shallow intensity profile gradients found by previous
    MIR diffraction limited imaging of MYSOs is indicative of outflow
    cavities (rather than alternative explanations such as rotational
    flattening).}

\item[--]{We show that the varied morphology observed can be
    attributed to outflow cavities seen at a variety of
    inclinations. This emphasises that outflows are an ubiquitous
    feature of massive star formation.}

\end{itemize}

\begin{acknowledgements}
  HEW acknowledges the financial support of the MPIfR. WJdW is grateful
  for the hospitality of G. Weigelt and the IR interferometry group at
  the MPIfR where this paper was finalised. J. Mottram is thanked for
  providing details of previous fits to the SEDs of many of the
  sources in this paper. This paper made use of information from the
  Red MSX Source survey database at www.ast.leeds.ac.uk/RMS which was
  constructed with support from the Science and Technology Facilities
  Council of the UK.
\end{acknowledgements}

\bibliographystyle{aa} 
\bibliography{bib}

\begin{thebibliography}{64}
\expandafter\ifx\csname natexlab\endcsname\relax\def\natexlab#1{#1}\fi

\bibitem[{{Beuther} {et~al.}(2002){Beuther}, {Schilke}, {Gueth}, {McCaughrean},
  {Andersen}, {Sridharan}, \& {Menten}}]{Beuther2002}
{Beuther}, H., {Schilke}, P., {Gueth}, F., {et~al.} 2002, A\&A, 387, 931

\bibitem[{{Bik} \& {Thi}(2004)}]{Bik2004}
{Bik}, A. \& {Thi}, W.~F. 2004, A\&A, 427, L13

\bibitem[{{Cesaroni} {et~al.}(2007){Cesaroni}, {Galli}, {Lodato}, {Walmsley},
  \& {Zhang}}]{Cesaroni2007}
{Cesaroni}, R., {Galli}, D., {Lodato}, G., {Walmsley}, C.~M., \& {Zhang}, Q.
  2007, in PPV, ed. {B.~Reipurth, D.~Jewitt, \& K.~Keil} (Tucson, AZ: Univ. of
  Arizona Press), 197--212

\bibitem[{{Cyganowski} {et~al.}(2009){Cyganowski}, {Brogan}, {Hunter}, \&
  {Churchwell}}]{Cyganowski2009}
{Cyganowski}, C.~J., {Brogan}, C.~L., {Hunter}, T.~R., \& {Churchwell}, E.
  2009, ApJ, 702, 1615

\bibitem[{{Davies} {et~al.}(2010){Davies}, {Lumsden}, {Hoare}, {Oudmaijer}, \&
  {de Wit}}]{Davies2010}
{Davies}, B., {Lumsden}, S.~L., {Hoare}, M.~G., {Oudmaijer}, R.~D., \& {de
  Wit}, W.-J. 2010, MNRAS, 402, 1504

\bibitem[{{De Buizer}(2006)}]{DeBuizer2006}
{De Buizer}, J.~M. 2006, ApJL, 642, L57

\bibitem[{{De Buizer}(2007)}]{DeBuizer2007}
{De Buizer}, J.~M. 2007, ApJL, 654, L147

\bibitem[{{De Buizer} {et~al.}(2000){De Buizer}, {Pi{\~n}a}, \&
  {Telesco}}]{DeBuizer2000}
{De Buizer}, J.~M., {Pi{\~n}a}, R.~K., \& {Telesco}, C.~M. 2000, ApJS, 130, 437

\bibitem[{{de Wit} {et~al.}(2010){de Wit}, {Hoare}, {Oudmaijer}, \&
  {Lumsden}}]{deWit2010}
{de Wit}, W.~J., {Hoare}, M.~G., {Oudmaijer}, R.~D., \& {Lumsden}, S.~L. 2010,
  A\&A, 515, A45

\bibitem[{{de Wit} {et~al.}(2007){de Wit}, {Hoare}, {Oudmaijer}, \&
  {Mottram}}]{Dewit2007}
{de Wit}, W.~J., {Hoare}, M.~G., {Oudmaijer}, R.~D., \& {Mottram}, J.~C. 2007,
  ApJL, 671, L169

\bibitem[{{de Wit} {et~al.}(2011){de Wit}, {Hoare}, {Oudmaijer},
  {N{\"u}rnberger}, {Wheelwright}, \& {Lumsden}}]{deWit2011}
{de Wit}, W.~J., {Hoare}, M.~G., {Oudmaijer}, R.~D., {et~al.} 2011, A\&A, 526,
  L5

\bibitem[{{DW09: de Wit} {et~al.}(2009){DW09: de Wit}, {Hoare}, {Fujiyoshi},
  {Oudmaijer}, {Honda}, {Kataza}, {Miyata}, {Okamoto}, {Onaka}, {Sako}, \&
  {Yamashita}}]{deWit2009}
{DW09: de Wit}, W.~J., {Hoare}, M.~G., {Fujiyoshi}, T., {et~al.} 2009, A\&A,
  494, 157

\bibitem[{{Egan} {et~al.}(2003){Egan}, {Price}, \& {Kraemer}}]{Egan2003}
{Egan}, M.~P., {Price}, S.~D., \& {Kraemer}, K.~E. 2003, in BAAS, Vol.~35, 1301

\bibitem[{{Follert} {et~al.}(2010){Follert}, {Linz}, {Stecklum}, {van Boekel},
  {Henning}, {Feldt}, {Herbst}, \& {Leinert}}]{Follert2010}
{Follert}, R., {Linz}, H., {Stecklum}, B., {et~al.} 2010, A\&A, 522, A17

\bibitem[{{Giannini} {et~al.}(2005){Giannini}, {Massi}, {Podio}, {Lorenzetti},
  {Nisini}, {Caratti o Garatti}, {Liseau}, {Lo Curto}, \&
  {Vitali}}]{Giannini2005}
{Giannini}, T., {Massi}, F., {Podio}, L., {et~al.} 2005, A\&A, 433, 941

\bibitem[{{Goddi} {et~al.}(2011{\natexlab{a}}){Goddi}, {Humphreys},
  {Greenhill}, {Chandler}, \& {Matthews}}]{Goddi2011}
{Goddi}, C., {Humphreys}, E.~M.~L., {Greenhill}, L.~J., {Chandler}, C.~J., \&
  {Matthews}, L.~D. 2011{\natexlab{a}}, ApJ, 728, 15

\bibitem[{{Goddi} {et~al.}(2011{\natexlab{b}}){Goddi}, {Moscadelli}, \&
  {Sanna}}]{Goddi_2011}
{Goddi}, C., {Moscadelli}, L., \& {Sanna}, A. 2011{\natexlab{b}}, A\&A, 535, L8

\bibitem[{{Grellmann} {et~al.}(2011){Grellmann}, {Ratzka}, {Kraus}, {Linz},
  {Preibisch}, \& {Weigelt}}]{Grellmann2011}
{Grellmann}, R., {Ratzka}, T., {Kraus}, S., {et~al.} 2011, A\&A, 532, A109

\bibitem[{{Guertler} {et~al.}(1991){Guertler}, {Henning}, {Kruegel}, \&
  {Chini}}]{Guertler1991}
{Guertler}, J., {Henning}, T., {Kruegel}, E., \& {Chini}, R. 1991, A\&A, 252,
  801

\bibitem[{{Guzm{\'a}n} {et~al.}(2010){Guzm{\'a}n}, {Garay}, \&
  {Brooks}}]{Guzman2010}
{Guzm{\'a}n}, A.~E., {Garay}, G., \& {Brooks}, K.~J. 2010, ApJ, 725, 734

\bibitem[{{Harvey} {et~al.}(1977){Harvey}, {Campbell}, \&
  {Hoffmann}}]{Harvey1977}
{Harvey}, P.~M., {Campbell}, M.~F., \& {Hoffmann}, W.~F. 1977, ApJ, 215, 151

\bibitem[{{Henning} {et~al.}(2000){Henning}, {Schreyer}, {Launhardt}, \&
  {Burkert}}]{Henning2000}
{Henning}, T., {Schreyer}, K., {Launhardt}, R., \& {Burkert}, A. 2000, A\&A,
  353, 211

\bibitem[{{Hoare}(2006)}]{Hoare2006}
{Hoare}, M.~G. 2006, ApJ, 649, 856

\bibitem[{{Hoare} {et~al.}(2007){Hoare}, {Kurtz}, {Lizano}, {Keto}, \&
  {Hofner}}]{Hoare2007PPV}
{Hoare}, M.~G., {Kurtz}, S.~E., {Lizano}, S., {Keto}, E., \& {Hofner}, P. 2007,
  in PPV, ed. {B.~Reipurth, D.~Jewitt, \& K.~Keil} (Tucson, AZ: Univ. of
  Arizona Press), 181--196

\bibitem[{{Jim{\'e}nez-Serra} {et~al.}(2007){Jim{\'e}nez-Serra},
  {Mart{\'{\i}}n-Pintado}, {Rodr{\'{\i}}guez-Franco}, {Chandler}, {Comito}, \&
  {Schilke}}]{Izas2007}
{Jim{\'e}nez-Serra}, I., {Mart{\'{\i}}n-Pintado}, J.,
  {Rodr{\'{\i}}guez-Franco}, A., {et~al.} 2007, ApJL, 661, L187

\bibitem[{{Kahn}(1974)}]{Kahn1974}
{Kahn}, F.~D. 1974, A\&A, 37, 149

\bibitem[{{Kataza} {et~al.}(2000){Kataza}, {Okamoto}, {Takubo}, {Onaka},
  {Sako}, {Nakamura}, {Miyata}, \& {Yamashita}}]{Kataza2000}
{Kataza}, H., {Okamoto}, Y., {Takubo}, S., {et~al.} 2000, in SPIE Conf. Ser.,
  ed. {M.~Iye \& A.~F.~Moorwood}, Vol. 4008, 1144

\bibitem[{{Kraus} {et~al.}(2010){Kraus}, {Hofmann}, {Menten}, {Schertl},
  {Weigelt}, {Wyrowski}, {Meilland}, {Perraut}, {Petrov}, {Robbe-Dubois},
  {Schilke}, \& {Testi}}]{Kraus2010}
{Kraus}, S., {Hofmann}, K., {Menten}, K.~M., {et~al.} 2010, Nature, 466, 339

\bibitem[{{Krumholz} {et~al.}(2009){Krumholz}, {Klein}, {McKee}, {Offner}, \&
  {Cunningham}}]{Krumholz2009}
{Krumholz}, M.~R., {Klein}, R.~I., {McKee}, C.~F., {Offner}, S.~S.~R., \&
  {Cunningham}, A.~J. 2009, Sci, 323, 754

\bibitem[{{Kuiper} {et~al.}(2010){Kuiper}, {Klahr}, {Beuther}, \&
  {Henning}}]{Kuiper2010}
{Kuiper}, R., {Klahr}, H., {Beuther}, H., \& {Henning}, T. 2010, ApJ, 722, 1556

\bibitem[{{Lagage} {et~al.}(2004){Lagage}, {Pel}, {Authier}, {Belorgey},
  {Claret}, {Doucet}, {Dubreuil}, {Durand}, {Elswijk}, {Girardot}, {K{\"a}ufl},
  {Kroes}, {Lortholary}, {Lussignol}, {Marchesi}, {Pantin}, {Peletier},
  {Pirard}, {Pragt}, {Rio}, {Schoenmaker}, {Siebenmorgen}, {Silber}, {Smette},
  {Sterzik}, \& {Veyssiere}}]{VISIR}
{Lagage}, P.~O., {Pel}, J.~W., {Authier}, M., {et~al.} 2004, The Messenger,
  117, 12

\bibitem[{{Larson} \& {Starrfield}(1971)}]{Larson1971}
{Larson}, R.~B. \& {Starrfield}, S. 1971, A\&A, 13, 190

\bibitem[{{Linz} {et~al.}(2009){Linz}, {Henning}, {Feldt}, {Pascucci}, {van
  Boekel}, {Men'shchikov}, {Stecklum}, {Chesneau}, {Ratzka}, {Quanz},
  {Leinert}, {Waters}, \& {Zinnecker}}]{Linz2009}
{Linz}, H., {Henning}, T., {Feldt}, M., {et~al.} 2009, A\&A, 505, 655

\bibitem[{{Lumsden} {et~al.}(2002){Lumsden}, {Hoare}, {Oudmaijer}, \&
  {Richards}}]{Lumsden2002}
{Lumsden}, S.~L., {Hoare}, M.~G., {Oudmaijer}, R.~D., \& {Richards}, D. 2002,
  MNRAS, 336, 621

\bibitem[{{Masqu{\'e}} {et~al.}(2011){Masqu{\'e}}, {Osorio}, {Girart},
  {Anglada}, {Garay}, {Estalella}, {Calvet}, \& {Beltr{\'a}n}}]{Masque2011}
{Masqu{\'e}}, J.~M., {Osorio}, M., {Girart}, J.~M., {et~al.} 2011, ApJ, 738, 43

\bibitem[{{Molinari} {et~al.}(1996){Molinari}, {Brand}, {Cesaroni}, \&
  {Palla}}]{Molinari1996}
{Molinari}, S., {Brand}, J., {Cesaroni}, R., \& {Palla}, F. 1996, A\&A, 308,
  573

\bibitem[{{Moscadelli} {et~al.}(2011){Moscadelli}, {Cesaroni}, {Rioja},
  {Dodson}, \& {Reid}}]{Moscadelli2011}
{Moscadelli}, L., {Cesaroni}, R., {Rioja}, M.~J., {Dodson}, R., \& {Reid},
  M.~J. 2011, A\&A, 526, A66

\bibitem[{{Mottram} {et~al.}(2007){Mottram}, {Hoare}, {Lumsden}, {Oudmaijer},
  {Urquhart}, {Sheret}, {Clarke}, \& {Allsopp}}]{Mottram2007}
{Mottram}, J.~C., {Hoare}, M.~G., {Lumsden}, S.~L., {et~al.} 2007, A\&A, 476,
  1019

\bibitem[{{Mottram} {et~al.}(2011){Mottram}, {Hoare}, {Urquhart}, {Lumsden},
  {Oudmaijer}, {Robitaille}, {Moore}, {Davies}, \& {Stead}}]{M2011}
{Mottram}, J.~C., {Hoare}, M.~G., {Urquhart}, J.~S., {et~al.} 2011, A\&A, 525,
  A149

\bibitem[{{Mueller} {et~al.}(2002){Mueller}, {Shirley}, {Evans}, \&
  {Jacobson}}]{Mueller2002}
{Mueller}, K.~E., {Shirley}, Y.~L., {Evans}, II, N.~J., \& {Jacobson}, H.~R.
  2002, ApJS, 143, 469

\bibitem[{{Palla} \& {Stahler}(1991)}]{Palla1991}
{Palla}, F. \& {Stahler}, S.~W. 1991, ApJ, 375, 288

\bibitem[{{Patel} {et~al.}(2005){Patel}, {Curiel}, {Sridharan}, {Zhang},
  {Hunter}, {Ho}, {Torrelles}, {Moran}, {G{\'o}mez}, \& {Anglada}}]{Patel2005}
{Patel}, N.~A., {Curiel}, S., {Sridharan}, T.~K., {et~al.} 2005, Nat, 437, 109

\bibitem[{{Porter} {et~al.}(1998){Porter}, {Drew}, \& {Lumsden}}]{Porter1998}
{Porter}, J.~M., {Drew}, J.~E., \& {Lumsden}, S.~L. 1998, A\&A, 332, 999

\bibitem[{{Reid} {et~al.}(2007){Reid}, {Menten}, {Greenhill}, \&
  {Chandler}}]{Reid2007}
{Reid}, M.~J., {Menten}, K.~M., {Greenhill}, L.~J., \& {Chandler}, C.~J. 2007,
  ApJ, 664, 950

\bibitem[{{Robitaille} {et~al.}(2007){Robitaille}, {Whitney}, {Indebetouw}, \&
  {Wood}}]{Robitaille2007}
{Robitaille}, T.~P., {Whitney}, B.~A., {Indebetouw}, R., \& {Wood}, K. 2007,
  ApJS, 169, 328

\bibitem[{{Rodriguez} \& {Bastian}(1994)}]{Rodriguez1994}
{Rodriguez}, L.~F. \& {Bastian}, T.~S. 1994, ApJ, 428, 324

\bibitem[{{Shakura} \& {Sunyaev}(1973)}]{Shakura1973}
{Shakura}, N.~I. \& {Sunyaev}, R.~A. 1973, A\&A, 24, 337

\bibitem[{{Sridharan} {et~al.}(2002){Sridharan}, {Beuther}, {Schilke},
  {Menten}, \& {Wyrowski}}]{Sridharan2002}
{Sridharan}, T.~K., {Beuther}, H., {Schilke}, P., {Menten}, K.~M., \&
  {Wyrowski}, F. 2002, ApJ, 566, 931

\bibitem[{{Terebey} {et~al.}(1984){Terebey}, {Shu}, \& {Cassen}}]{Terebey1984}
{Terebey}, S., {Shu}, F.~H., \& {Cassen}, P. 1984, ApJ, 286, 529

\bibitem[{{Ulrich}(1976)}]{Ulrich1976}
{Ulrich}, R.~K. 1976, ApJ, 210, 377

\bibitem[{{Urquhart} {et~al.}(2007){Urquhart}, {Busfield}, {Hoare}, {Lumsden},
  {Clarke}, {Moore}, {Mottram}, \& {Oudmaijer}}]{Urquhart2007a}
{Urquhart}, J.~S., {Busfield}, A.~L., {Hoare}, M.~G., {et~al.} 2007, A\&A, 461,
  11

\bibitem[{{Urquhart} {et~al.}(2008){Urquhart}, {Hoare}, {Lumsden}, {Oudmaijer},
  \& {Moore}}]{James-RMS}
{Urquhart}, J.~S., {Hoare}, M.~G., {Lumsden}, S.~L., {Oudmaijer}, R.~D., \&
  {Moore}, T.~J.~T. 2008, in ASP Conf. Ser., Vol. 387, Massive Star Formation:
  Observations Confront Theory, ed. {H.~Beuther, H.~Linz, \& T.~Henning} (San
  Fransico, CA: ASP), 381

\bibitem[{{Urquhart} {et~al.}(2011){Urquhart}, {Moore}, {Hoare}, {Lumsden},
  {Oudmaijer}, {Rathborne}, {Mottram}, {Davies}, \& {Stead}}]{Urquhart2011}
{Urquhart}, J.~S., {Moore}, T.~J.~T., {Hoare}, M.~G., {et~al.} 2011, MNRAS,
  410, 1237

\bibitem[{{van der Tak} {et~al.}(2000){van der Tak}, {van Dishoeck}, {Evans},
  \& {Blake}}]{vdt2000}
{van der Tak}, F.~F.~S., {van Dishoeck}, E.~F., {Evans}, II, N.~J., \& {Blake},
  G.~A. 2000, ApJ, 537, 283

\bibitem[{{Vehoff} {et~al.}(2010){Vehoff}, {Hummel}, {Monnier}, {Tuthill},
  {N{\"u}rnberger}, {Siebenmorgen}, {Chesneau}, \& {Duschl}}]{Vehoff2010}
{Vehoff}, S., {Hummel}, C.~A., {Monnier}, J.~D., {et~al.} 2010, A\&A, 520, A78

\bibitem[{{Vinkovi{\'c}} \& {Jurki{\'c}}(2007)}]{Vinkovic2007}
{Vinkovi{\'c}}, D. \& {Jurki{\'c}}, T. 2007, ApJ, 658, 462

\bibitem[{{Walsh} {et~al.}(1997){Walsh}, {Hyland}, {Robinson}, \&
  {Burton}}]{Walsh1997}
{Walsh}, A.~J., {Hyland}, A.~R., {Robinson}, G., \& {Burton}, M.~G. 1997,
  MNRAS, 291, 261

\bibitem[{{Wheelwright} {et~al.}(2010){Wheelwright}, {Oudmaijer}, {de Wit},
  {Hoare}, {Lumsden}, \& {Urquhart}}]{MeCO}
{Wheelwright}, H.~E., {Oudmaijer}, R.~D., {de Wit}, W.~J., {et~al.} 2010,
  MNRAS, 408, 1840

\bibitem[{{Whitney} {et~al.}(2003{\natexlab{a}}){Whitney}, {Wood}, {Bjorkman},
  \& {Cohen}}]{W22003}
{Whitney}, B.~A., {Wood}, K., {Bjorkman}, J.~E., \& {Cohen}, M.
  2003{\natexlab{a}}, ApJ, 598, 1079

\bibitem[{{Whitney} {et~al.}(2003{\natexlab{b}}){Whitney}, {Wood}, {Bjorkman},
  \& {Wolff}}]{W12003}
{Whitney}, B.~A., {Wood}, K., {Bjorkman}, J.~E., \& {Wolff}, M.~J.
  2003{\natexlab{b}}, ApJ, 591, 1049

\bibitem[{{Wolfire} \& {Cassinelli}(1987)}]{Wolfire1987}
{Wolfire}, M.~G. \& {Cassinelli}, J.~P. 1987, ApJ, 319, 850

\bibitem[{{Yorke} \& {Sonnhalter}(2002)}]{YorkeandSonnhalter2002}
{Yorke}, H.~W. \& {Sonnhalter}, C. 2002, ApJ, 569, 846

\bibitem[{{Zhang} \& {Tan}(2011)}]{Zhang2011}
{Zhang}, Y. \& {Tan}, J.~C. 2011, ApJ, 733, 55

\bibitem[{Zinnecker \& Yorke(2007)}]{ZinneckerandYorke2007}
Zinnecker, H. \& Yorke, H. 2007, ARA\&A, 45, 481

\end{thebibliography}

\appendix

\section{Notes on individual objects}

\setcounter{secnumdepth}{1}

\label{notes}

Here we discuss the images and modelling results for each resolved MYSO.

\subsection{A: G263.7759$-$00.4281}
The 20~$\mu$m morphology of G263.7759$-$00.4281 (IRAS 08448-4343) is
reminiscent of the cavities of a bipolar outflow seen close to edge
on. The direction of the elongation seen in the MIR image is
consistent with that seen in the NIR via 2MASS images. The radio
luminosity of the object is consistent with a jet rather than an
H{\sc{ii}} region \citep{Hoare2007PPV}. This supports the notion that
the bulk of the MIR emission arises in the walls of cavities evacuated
by an outflow. Furthermore, the extension in the NW/SE direction is
aligned with the $\mathrm{H_2}$ jet detected by \citet{Giannini2005},
confirming that this object drives an outflow and the MIR morphology
is associated with outflow activity.

\smallskip

The SED of this object can be reproduced with the selected model with
an inclination of $i=87$, which is consistent with the notion that
this object is seen close to edge on. The intensity profile of the
model is not entirely consistent with the data. However, the
axis-symmetric model used cannot account for the slight asymmetry of
the image. Indeed, the model can approximately recreate the extension
of the southern part of the observed bipolar structure but not the
northern half (see Fig. \ref{rad_profs}, panel A). The model cannot
reproduce asymmetry at these wavelengths in an edge on
configuration. Therefore, we surmise that the model reproduces the
data as well as can be expected.

\subsection{B: G265.1438+01.4548}
This object exhibits a fairly symmetric morphology in the MIR. If the
generic model featuring outflows is correct, this would suggest that
this object is seen at a relatively low inclination. Modelling of this
object's CO bandhead emission is consistent with this scenario
\citep{Bik2004,MeCO}. The SED and intensity profile are relatively
well reproduced by the generic model with an inclination of $i=32$,
which is consistent with the hypothesis that this object is viewed at
a low inclination. The model slightly under-predicts the flux at
distances greater than $\sim$2\arcsec$\,$from the centre of the intensity
distribution. However, examination of the image reveals that this may
not be related to the source as the core of the emission is
concentrated within 2\arcsec. Therefore, this slight discrepancy is
neglected.

\subsection{C: G268.3957$-$00.4842}
G268.3957$-$00.4842 exhibits a relatively symmetric morphology in the
MIR. As before, it is suggested that this is the result of a
relatively low inclination. Indeed, the object's SED and intensity
profile are reproduced by a model with an inclination of $i=30$,
consistent with the hypothesis that this object is observed at a low
inclination.

\subsection{D: G269.1586$-$01.1383}
The 20~$\mu$m image of G269.1586$-$01.1383 incorporates an additional,
localised source of flux approximately 7\arcsec$\,$to the NE. Both
sources exhibit an extended, slightly cometary morphology reminiscent of a
H{\sc{ii}} region. However, radio observations indicate that while the
northern source is a H{\sc{ii}} region \citep{Urquhart2007a}, the
southern source is likely to be a YSO. This is confirmed via low
resolution NIR spectra. The spectrum of the southern source exhibits
$\mathrm{H_2}$ line emission, suggesting that it drives an outflow.

\smallskip

The complex appearance of this object is difficult to recreate with
the 2D, axis-symmetric codes of Whitney et al.. In principle, a bipolar
cometary morphology could be recreated by warm dust emission from the walls of cavities
carved by a bipolar outflow with a large opening angle. However, the
image contains no hint of a bipolar structure. Therefore, it might be
expected that the model cannot reproduce the intensity profile of this
source, as is found to be the case. We note that this object exhibits
a notably different morphology to the rest of the sample, which may
indicate it represents a different phase of massive star formation
than the other MYSOs. This is partially substantiated by the fact that
this object is associated with a H{\sc{ii}} region.

\subsection{E: G310.0135+00.3892}

G310.0135+00.3892 (IRAS 13481-6124) was observed with the VLTI and
AMBER by \citet{Kraus2010}. These authors reconstructed images of this
object in the $K$-band and detected a disc seen under moderate
inclination.  In addition, \citet{Kraus2010} discovered signs of an
outflow orientated in the NE/SW direction, perpendicularly to the
disc. The 20~$\mu$m morphology of this object is slightly extended in the NE/SW direction, i.e. along the direction of the outflow. This is consistent with the notion that the majority of the MIR emission traces warm dust in the walls of outflow cavities.

\smallskip

The observed SED and intensity profile are reproduced relatively well
by a model with an inclination of $i=32$, close to the intermediate
inclination derived by \citet{Kraus2010}. The model under-predicts the
flux at distances greater than $\sim$1.5\arcsec$\,$from the peak of the
intensity distribution. However, examination of the image reveals that
at this distance the morphology is not symmetric. The axis-symmetric
model cannot reproduce this morphology. Therefore, it is surmised that
the model reproduces the data as well as possible and this
slight discrepancy is neglected.

\subsection{F: G318.0489+00.0854} 
The 20~$\mu$m image of G318.0489+00.0854 exhibits a slightly cometary
morphology. This object is associated with two H{\sc{ii}} regions
detected via radio observations, both within several arcseconds
\citep{Urquhart2007a}. The near/far ambiguity over the distance to this
source has yet to be resolved. As a result, it is impossible to
unambiguously model its intensity profile and SED. Therefore, we do not
attempt to reproduce the observations of this object.

\subsection{G: G332.2941+02.2799} 

G332.2941+02.2799 (IRAS 16019-4903) exhibits a significantly extended
morphology in the NE/SW direction. This is also seen in the NIR in
2MASS images and at 10~$\mu$m \citep[see][]{Mottram2007}. This object
is known to be associated with several bipolar outflows and the
direction of the most prominent outflow is aligned with the extension
seen in the infrared images \citep{Henning2000}. As a result, the
observed morphology is consistent with the notion that the MIR flux
largely arises from warm dust in outflow cavity walls.

\smallskip

Since these observations were taken, it has been determined that the
luminosity of G332.2941+02.2799 is 844~$L_{\odot}$
\citep[see][]{M2011}. This places the spectral type of this object at
approximately B2/B3. Therefore, this object is an intermediate mass
YSO rather than a massive YSO and should be excluded from the
sample. Nonetheless, we attempt to recreate the observations with the
generic model developed for the sample of MYSOs. The model cannot
fully recreate the extension of the intensity profile observed, even
with a high inclination and stellar luminosity. This may imply that
the model developed for MYSOs is not directly applicable to
intermediate mass YSOs. Alternatively, the strong outflow may cause
shock heating in the cavities, which is not included in the model.

\subsection{H: G332.9868$-$00.4871}
G332.9868$-$00.4871 exhibits a fairly symmetric morphology. As before,
it is suggested that this is the result of a relatively low
inclination. The object's SED is well reproduced by a model with an
inclination of $i=15\degr$, as is the object's intensity profile. The
model intensity profile deviates from that observed at a distance of
$\sim$1.75\arcsec from the centre. However, this is where the object's
flux distribution falls to the level of the background and thus this
discrepancy is not considered significant.

\subsection{I: G339.6221$-$00.1209}

This object exhibits a slightly cometary morphology. As discussed in
the case of G269.1586$-$01.1383, this is difficult to recreate at
20~$\mu$m with an axis-symmetric code. We note that this object is the
most luminous in the sample with $L=5.2 \times 10^5
L_{\odot}$. \citet{M2011} find that there is not a significant
population of radio quiet MYSOs with such a luminosity. Sources with
luminosities greater than $L\sim10^{5}L_{\odot}$ are almost inevitably
associated with a H{\sc{ii}} region. Therefore, it might be expected
that this object is associated with a H{\sc{ii}} region. While this
object is a radio non-detection \citep{Urquhart2007a}, it is possible
that a weak H{\sc{ii}} region has escaped detection due to the large
distance to this object (13~kpc). This could explain the cometary
morphology of this source.

\smallskip

A relatively inclined model ($i=60\degr$) reproduces the object's SED
and intensity profile relatively well. The mm flux is under predicted,
but at the large distance to this object, several sources could have
been included in the mm beam, resulting in an over-estimation of the
flux. We note that, although the model recreates the observed
intensity profile, the model image exhibits bi-polar structure while
the observed morphology is appears more mono-polar. As discussed above, this source
may harbour a H{\sc{ii}} region, which could explain this discrepancy.

\subsection{J: G343.5024$-$00.0145}
The image of G343.5024$-$00.0145 exhibits a complex, notably extended
morphology. Extended nebulosity towards the SE is also visible in
2MASS and Glimpse IRAC images. The extension of the 20~$\mu$m image is
broadly consistent with that seen at shorter wavelengths, although
these data exhibit a more complex morphology. This source was detected
at 4.8~GHz \citep{Urquhart2007a}, and is thus classified as a
H{\sc{ii}} in the RMS catalogue. Therefore, it might be expected that
the model developed to recreate the observations of radio quiet MYSOs
cannot reproduce the complex morphology exhibited by this
object. Indeed, the model that recreates the majority of the object's
SED fails to match the spatial extension of the 20~$\mu$m flux. We
suggest that this is likely due to some of the observed flux
originating in dust surrounding a H{\sc{ii}} region, rather than the
central source.

\subsection{K: G343.5213$-$00.5171}
The image of G343.5213$-$00.5171 reveals three localised sources. The
central source clearly dominates the flux. However, one of the
secondary sources is within 2\arcsec of the primary source. It is
apparent that secondary flux will contaminate the intensity profile of
the primary. Indeed, the image reveals that this source is less
centrally concentrated than its isolated counterparts. A limited range
of angles was used in constructing the azimuthally averaged intensity
profile of this object in an attempt to limit contamination, but it is
not clear how effective this was. The model fails to simultaneously
reproduce the broad central maximum of the image and the observed
SED. We suggest that this is the result of contamination of the
intensity profile by flux from the neighbouring secondary source.

\subsection{L: G345.0061+01.7944}
G345.0061+01.7944 exhibits an extended morphology similar to that of
G332.2941+02.2799. This suggests that the MIR image traces outflow
structure seen close to edge-on. Indeed, the NIR spectrum of this
object exhibits $\mathrm{H_2}$ emission indicative of outflow
activity. Unlike the case of G332.2941+02.2799, a satisfactory fit to
both the SED and the intensity profile was found. This is consistent
with the hypothesis that the difficulties in reproducing the
observations of G332.2941+02.2799 were the result of applying a model
developed for high luminosity MYSOs to the case of an intermediate
mass YSO.

\subsection{M: G349.7215+00.1203} 
G349.7215+00.1203 is the second most luminous object in the sample
with $L=3.1 \times 10^{5}L_{\odot}$. The image of this object
reveals a cometary shaped source of flux to the NW of the primary
object. Radio observations confirm that the additional source of flux
is a H{\sc{ii}} region \citep{Urquhart2007a}. The model reproduces
most of the features of the SED and intensity profile of this
object. The model under-predicts the flux at
$\sim$100~$\mu$m. However, this could be due to the measured flux
including a contribution from the H{\sc{ii}} region. The model
intensity profile exhibits a slightly different slope than that
observed, although it is essentially within the uncertainty. We
suggest that this could also be due to the presence of the H{\sc{ii}}
region.

\end{document}